\documentclass[a4paper,11pt]{article}
\usepackage{amsmath,amssymb,mathtools}
\usepackage{amsthm}
\usepackage{bbold}
\usepackage{graphicx}
\usepackage{epic}
\usepackage{hyperref}
\usepackage{cite}
\usepackage{multicol,color,longtable}

\definecolor{darkred}{rgb}{0.65,0.15,0}
\hypersetup{pdfborder={0 0 0},colorlinks=true,urlcolor=darkred,citecolor=blue,linkcolor=darkred,linktocpage=true}

\DeclareMathOperator{\Z}{\mathbb{Z}}
\DeclareMathOperator{\R}{\mathbb{R}}
\DeclareMathOperator{\C}{\mathbb{C}}
\DeclareMathOperator{\id}{\mathbb{1}}

\DeclareMathOperator{\g}{\mathfrak{g}}
\DeclareMathOperator{\h}{\mathfrak{h}}
\DeclareMathOperator{\n}{\mathfrak{n}}
\DeclareMathOperator{\q}{\mathfrak{q}}
\DeclareMathOperator{\qh}{\mathfrak{q}_{1/2}}
\DeclareMathOperator{\ad}{\mathrm{ad}}
\DeclareMathOperator{\Aut}{\mathrm{Aut}}
\DeclareMathOperator{\im}{\mathrm{Im}}

\DeclareMathOperator{\diag}{\mathrm{diag}}
\DeclareMathOperator{\Sl}{\mathfrak{sl}}
\DeclareMathOperator{\so}{\mathfrak{so}}
\DeclareMathOperator{\su}{\mathfrak{su}}
\DeclareMathOperator{\usp}{\mathfrak{usp}}
\DeclareMathOperator{\SL}{SL}
\DeclareMathOperator{\SO}{SO}
\DeclareMathOperator{\SU}{SU}
\DeclareMathOperator{\U}{U}
\DeclareMathOperator{\Spin}{Spin}

\newcommand{\e}{E_{10}}
\newcommand{\ke}{K(E_{10})}
\newcommand{\de}{DE_{10}}
\newcommand{\kde}{K(DE_{10})}
\newcommand{\be}{BE_{10}}
\newcommand{\kbe}{K(BE_{10})}
\newcommand{\aethree}{AE_3}
\newcommand{\kaethree}{K(AE_3)}
\newcommand{\aed}{AE_d}
\newcommand{\kaed}{K(AE_d)}
\newcommand{\gtwo}{G_2^{++}}
\newcommand{\kgtwo}{K(G_2^{++})}
\newcommand{\ind}{\mathtt}

\newcommand{\lb}{\left[}
\newcommand{\rb}{\right]}

\theoremstyle{plain}
\newtheorem*{theorem}{Theorem}
\theoremstyle{plain}
\newtheorem*{corollary}{Corollary}

\begin{document}

\renewcommand\footnoterule{}

\thispagestyle{empty}
\mbox{ }
\vspace{20mm}

\renewcommand{\thefootnote}{\fnsymbol{footnote}}

\begin{center}
{\LARGE \bf On spinorial representations of involutory \\[3mm]  
    subalgebras of  Kac--Moody algebras\footnote{Based on
    lectures given by H.~Nicolai at the School {\it Partition Functions and Automorphic
    Forms},\\[0.5mm] BLTP JINR, Dubna, Russia, 28 January - 2 February 2018.}}\\[10mm]

\vspace{8mm}
\normalsize
{\large  Axel Kleinschmidt${}^{1,2}$, Hermann Nicolai${}^1$ and Adriano Vigan\`o${}^1$}

\vspace{10mm}
${}^1${\it Max-Planck-Institut f\"ur Gravitationsphysik (Albert-Einstein-Institut)\\
Am M\"{u}hlenberg 1, 14476 Potsdam, Germany}
\vskip 1 em
${}^2${\it International Solvay Institutes\\
ULB Campus Plaine CP231, 1050 Brussels, Belgium}

\vspace{20mm}

\hrule

\vspace{10mm}

\begin{tabular}{p{12cm}}
{\small
The representation theory of involutory (or `maximal compact')
subalgebras of infinite-dimensional Kac--Moody 
algebras is largely {\em terra incognita}, especially with regard to fermionic
(double-valued) representations. Nevertheless, certain distinguished such representations
feature prominently in proposals of possible symmetries underlying M theory, both at
the classical and the quantum level. Here we summarise recent efforts to study \textit{spinorial} representations systematically, most notably for the case of the hyperbolic Kac--Moody algebra $E_{10}$ where spinors of the involutory subalgebra $K(E_{10})$ are expected to play a role in describing algebraically the fermionic sector of $D=11$ supergravity and M theory.
Although these results remain very incomplete, they also point towards the beginning
of a possible  explanation of the fermion structure observed in the Standard Model 
of Particle Physics.}
\end{tabular}

\vspace{5mm}
\hrule

\end{center}

\newpage

\tableofcontents

\renewcommand{\thefootnote}{\arabic{footnote}}
\setcounter{footnote}{0}

\renewcommand\footnoterule{\kern-3pt \hrule width 2in \kern 2.6pt}

\section{Introduction}

Fermionic fields are central in all fundamental particle models as constituents of matter. They are characterised as being double-valued representations of the space-time symmetry group, notably the spin cover of the Lorentz group $\Spin(1,D-1)$ in $D$ space-time dimensions. For instance, the Standard Model of particle physics contains $48$ spin-$\frac12$ particles, if one includes also right-handed neutrinos in the counting, and these transform also under the gauge symmetries SU(3)$_c \,\times$ SU(2)$_w \, \times$ U(1)$_Y$. To date there is no satisfactory explanation 
of this precise set of spin-$\frac12$ particles of the Standard Model of particle physics. 
Fermionic fields are also essential and inevitable in all supersymmetric theories, notably supergravity and superstring theory. Investigations of the bosonic sectors of these theories have led to conjectured gigantic symmetry groups governing the bosonic dynamics. The most well-known of these conjectures involve the infinite-dimensional Kac--Moody groups $E_{10}$ and $E_{11}$~\cite{Julia:1980gr,West:2001as,Damour:2002cu}. Implicit in these conjectures is that there are also infinite-dimensional extensions of spin groups acting on the fermionic fields. At the Lie algebra level, these infinite-dimensional extensions are given by fixed points of an involution acting on the Kac--Moody algebra and are known by the name $K(E_{10})$ and $K(E_{11})$ and often called maximal compact subalgebras.\footnote{Here and in the remainder of these notes, we abuse notation by using $E_n$ to denote both the Lie algebra and the associated group.} 

In order to understand the fermions in the context of these conjectures it is therefore important to understand the fermionic (double-valued) representations of the maximal compact subalgebras $K(E_n)$. For $n\geq 9$,  $K(E_n)$ does not belong to any standard class of Lie algebras, and there is
no off-the-shelf representation theory available for finding the fermionic representations (whereas
for finite-dimensional $E_n$ groups, {\em i.e.}~$n\leq 8$, the maximal compact subalgebras
{\em are} of standard type). 
By analysing the structure of maximal supergravity, some first spin-$\frac12$ and spin-$\frac32$ representations of $K(E_{10})$ were found in~\cite{deBuyl:2005zy,Damour:2005zs,deBuyl:2005sch,Damour:2006xu}. The fermionic representations found there also have the unusual property of unifying the distinct type IIA and type IIB spinors of $D=10$ maximal supersymmetry~\cite{Kleinschmidt:2006tm,Kleinschmidt:2007zd}. Moreover, as we shall also review in these proceedings, they hold the potential of explaining the fermionic content of the Standard Model of particle physics~\cite{Meissner:2014joa,Kleinschmidt:2015sfa,Meissner:2018gtx}.

The purpose of this article is to summarise and systematise more recent advances on the representation theory of $K(E_{10})$ and similar (hyperbolic) Kac--Moody algebras~\cite{Kleinschmidt:2013eka,Kleinschmidt:2014uwa,Kleinschmidt:2016ivr}. Specifically, we shall explain:
\begin{itemize}
\item the construction of involutory subalgebras of Kac--Moody algebras;
\item the basic conditions, called Berman relations, that any representation has to satisfy;
\item the construction of spin-$\frac12$ representations for simply-laced algebras;
\item the extension up to spin-$\frac72$ for the simply-laced $E_{10}$, $DE_{10}$ and 
         $AE_d$ family (for $d\geq 3$);
\item the extension to higher spin for the non-simply-laced $AE_3$, $G_2^{++}$ and $BE_{10}$;
\item the properties of these representations, in particular the fact that they are unfaithful and imply the existence of finite-dimensional quotients of the infinite-dimensional involutory algebras,
as we will illustrate with several examples;
\item the connection between $K(E_{10})$ and the fermions of the Standard Model.
\end{itemize}
Other work on fermionic representations includes~\cite{West:2003fc,Nicolai:2004nv,Kleinschmidt:2006dy,Damour:2009zc,Hainke:2011mt,Damour:2013eua,Damour:2014cba,Ghatei:2017,Lautenbacher:2017}. Let us stress again that these results represent only very
first steps, and  that many open problems remain which will probably require completely
new insights and methods.

\section{Involutory subalgebras of Kac--Moody algebras}

We review some basic facts about Kac--Moody algebras and their maximal compact subalgebras, referring the reader to~\cite{Kac:1990gs} for proofs and details.

\subsection{Basic definitions}

An $n\times n$ matrix $A$ which satisfies
\begin{subequations}
\begin{align}
& A_{ii} = 2 \quad \forall i = 1, \dotsc, n , \\
& A_{ij} \in \Z_{\leq 0} \quad (i \neq j) , \\
& A_{ij} = 0 \iff A_{ji} = 0 ,
\end{align}
\end{subequations}
where $\Z_{\leq 0}$ denotes the non-positive integers, is called a \emph{generalised Cartan matrix} (we will frequently refer to $A$ simply as the Cartan matrix).
We always assume $\det A\ne 0$.

The \emph{Kac--Moody algebra} $\g\equiv\g(A)$ is the Lie algebra generated 
by the $3n$ generators $\{e_i,f_i,h_i\}$ subject to the relations
\begin{subequations}
\label{chev-serre}
\begin{align}
[h_i, h_j] & = 0 , \\
\label{chevalley2}
[h_i, e_j] & = A_{ij} e_j , \\
[h_i, f_j] & = -A_{ij} f_j , \\
[e_i, f_j] & = \delta_{ij} h_j , \\
\label{serre1}
\ad_{e_i}^{1-A_{ij}} e_j & = 0 , \\
\label{serre2}
\ad_{f_i}^{1-A_{ij}} f_j & = 0 .
\end{align}
\end{subequations}
Each triple $\{e_i,f_i,h_i\}$ generates a subalgebra isomorphic to $\Sl(2,\C)$, as 
is evident from relations~\eqref{chev-serre}. This construction of the algebra in terms
of generators and relations  is generally referred to as the \emph{Chevalley-Serre presentation}.

A generalised Cartan matrix $A$ is called \emph{simply-laced} if the off-diagonal 
entries of $A$ are either $0$ or $-1$. $A$ is called \emph{symmetrizable} if there exists 
a diagonal matrix $D$ such that $DA$ is symmetric.
We will say that $\g$ is simply-laced (symmetrizable) if its generalised Cartan matrix $A$ is simply-laced (symmetrizable). Analogously, the algebra is \emph{symmetric} if its Cartan matrix is.
In the finite dimensional case an algebra is simply-laced if and only if it is symmetric, 
but this is not true in the general infinite-dimensional case. If the Cartan matrix $A$ is indefinite,
the associated (infinite-dimensional) Kac--Moody algebra is called {\em indefinite}; it is 
called {\em hyperbolic} if the removal of any node in the Dynkin diagram leaves an
algebra that is either of finite or affine type; with this extra requirement, the rank is bounded
above by $r\leq 10$, with four such maximal rank algebras, 
$E_{10}$, $DE_{10}$, $BE_{10}$ and $CE_{10}$~\cite{Kac:1990gs}. Our interest here is mainly with hyperbolic algebras, as these appear to be 
the relevant ones for M theory. 

The \emph{Dynkin diagram} of a Kac--Moody algebra $\g$ is a set of 
vertices and edges built with the following rules:
\begin{itemize}
\item for each $i=1,\dotsc,n$ there is an associated node in the diagram;
\item if $|A_{ij}|=1$ or $|A_{ji}|=1$, the are $\max(|A_{ij}|,|A_{ji}|)$ lines between the nodes $i$ and $j$ and an arrow from $j$ to $i$ if $|A_{ij}|>|A_{ji}|$. If both $|A_{ij}|>1$ and $|A_{ji}|>1$, the line from $i$ to $j$ is decorated with the pair $(|A_{ij}|, |A_{ji}|)$.
\end{itemize}
Hence, one can read the Cartan matrix from the Dynkin diagram and \emph{vice versa}.

Let $\h\coloneqq\text{span}(h_1,\dotsc,h_n)$ be the $n$-dimensional algebra of semi-simple elements, called the \emph{Cartan subalgebra}. Furthermore, we let $\n_+$ be the quotient of the free Lie algebra generated by $\{e_1,\dotsc,e_n\}$ modulo the Serre relation~\eqref{serre1} and, similarly, $\n_-$ is the quotient of the free Lie algebra generated by $\{f_1,\dotsc,f_n\}$ modulo the Serre relation~\eqref{serre2}. Then we have the triangular vector space decomposition
\begin{equation}
\g = \n_+ \oplus \h \oplus \n_- .
\end{equation}
All three spaces are Lie subalgebras of $\g$, but the sum is not direct as a sum of Lie algebras as for example $\lb \h , \n_{\pm}\rb \subset \n_{\pm}$.

The number of the nested commutators which generate $\n_+$ and $\n_-$ is a priori infinite, but the Serre relations~\eqref{serre1} and~\eqref{serre2} impose non-trivial relations among them; this makes the algebra finite or infinite-dimensional, according to whether the generalised Cartan matrix 
is positive definite or not.

The adjoint action of the Cartan subalgebra $\h$ on $\n_+$ and $\n_-$ is diagonal, which means that it is possible to decompose $\g$ into eigenspaces
\begin{equation}
\g_\alpha \coloneqq \{x \in \g \mid [h,x] = \alpha(h)x \quad \forall h \in \h \} ,
\end{equation}
where $\alpha:\h\to\C$ belongs to the dual space of the Cartan subalgebra.
The eigenspaces $\g_\alpha$ are called \emph{root spaces} and the linear functionals $\alpha\in\h^*$ \emph{roots}.
The \emph{multiplicity} of a root $\alpha$ is defined as the dimension of the related root space $\g_\alpha$.
Relation~\eqref{chevalley2} implies that each $e_j$ is contained in an eigenspace $\g_{\alpha_j}$, whose corresponding root $\alpha_j$ is called a \emph{simple root}.
Accordingly, $f_j$ belongs to $\g_{-\alpha_j}$ with simple root $-\alpha_j$.

The lattice $Q\coloneqq\bigoplus_{i=1}^n\Z\alpha_i$ provides a grading of $\g$
\begin{equation}
[\g_\alpha,\g_\beta] \subseteq \g_{\alpha+\beta} ,
\end{equation}
together with the root space decomposition
\begin{equation}
\g = \bigoplus_{\alpha\in Q} \g_\alpha .
\end{equation}
Any root of $\g$ is a non-negative or non-positive integer linear combination of the simple roots $\alpha_i$.

Let us assume that $\g$ is symmetrizable; this allows to define a bilinear form on 
the Cartan subalgebra $\h$, which can be extended to an invariant bilinear form on the whole algebra $\g$ obtaining the so-called \emph{(generalised) Cartan--Killing form}.
We are more interested in the scalar product $(\cdot\mid\cdot)$ among the roots, \emph{i.e.}~elements of $\h^*$.
For any pair of simple roots $\alpha_i,\alpha_j$ it is given by
\begin{equation}
\label{symC}
(\alpha_i\mid\alpha_j) = B_{ij} ,
\end{equation}
where $B$ is the symmetrization of the Cartan matrix $A$.
This defines the scalar product between two generic roots, since any root can be written as a linear combination of simple roots.

The scalar product gives a natural notion of root length, \emph{i.e.}~$\alpha^2\coloneqq(\alpha \,|\, \alpha)$.
In the finite dimensional case the roots are all spacelike, meaning $\alpha^2>0$;
this reflects the fact that the scalar product has Euclidean signature.
In the infinite-dimensional case this is no longer true:
the spacelike roots are called \emph{real roots}, the non-spacelike ones $\alpha^2\leq0$ are called \emph{imaginary roots}.
In particular, the lightlike roots $\alpha^2=0$ are called \emph{null roots}.

The real roots are non-degenerate (\emph{i.e.}~the corresponding root spaces are one dimensional) 
as in the finite dimensional case, while the imaginary roots are not.
The degeneracy of the imaginary roots is still an open problem, since a general 
closed formula for the multiplicity of an imaginary root for non-affine $\g$ is yet to 
come.\footnote{In fact, the multiplicities are not even known in closed form for the simplest such algebra with Cartan matrix
\begin{equation*}
A = \begin{pmatrix} 2 & -3\\ -3 & 2 \end{pmatrix} \,.
\end{equation*}}

The discussion above was phrased in terms of complex Lie algebras. However, as is evident from~\eqref{chev-serre}, there is a natural basis in which the structure constants are real and therefore we can also consider $\g$ over the real numbers in what is called the \emph{split form} by simply taking the same generators and only real combinations of their multi-commutators. This split real Kac--Moody algebra will be denoted by $\g$ in the following and from now on all discussions are for real Lie algebras. Furthermore we restrict attention to 
symmetrizable Cartan matrices.

\subsection{Maximal compact subalgebras}
\label{sec:K}

We introduce the \emph{Cartan--Chevalley involution} $\omega \in \Aut(\g)$, defined by
\begin{equation}\label{CCinvolution}
\omega(e_i) = -f_i, \quad
\omega(f_i) = -e_i, \quad
\omega(h_i) = -h_i ;
\end{equation}
its action on a root space is therefore $\omega(\g_\alpha) = \g_{-\alpha}$.
Moreover it is linear and satisfies
\begin{equation}
\omega(\omega(x)) = x , \quad
\omega([x,y]) = [\omega(x),\omega(y)] .
\end{equation}
The latter relation allows to extend the involution from (\ref{CCinvolution}) to the whole algebra.
The involutory, or \emph{maximal compact} subalgebra, $K(\g)$ of $\g$ is the set of fixed points 
under the Cartan-Chevalley involution
\begin{equation}
K(\g) \coloneqq \{x \in \g \mid \omega(x) = x\} .
\end{equation}
The name arises from the analogy with the finite dimensional split real case, where 
for example $K(\Sl(n,\R))=\so(n,\R)$: here $\so(n,\R)$ is the Lie algebra of the 
maximal compact subgroup $\SO(n,\R)$ of $\SL(n,\R)$ and the involution is simply
$\omega(x) = - x^T$.

\begin{figure}
\centering
\begin{picture}(20,50)
\put(0,0){\circle{10}}
\put(5,0){\line(1,0){20}}
\put(30,0){\circle{10}}
\put(35,-1){\line(1,0){20}}
\put(35,1){\line(1,0){20}}
\put(60,0){\circle{10}}
\drawline(35,0)(40,5)
\drawline(35,0)(40,-5)
\drawline(55,0)(50,5)
\drawline(55,0)(50,-5)
\put(-2,10){$1$}
\put(28,10){$2$}
\put(57,10){$3$}
\end{picture}
\caption{Dynkin diagram of $\aethree$}
\label{dynkin:aethree}
\end{figure}
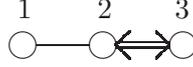

The combination $x_i\coloneqq e_i-f_i$ is the only one which is invariant under $\omega$ in each triple (corresponding to the compact $\so(2,\R)\subset\Sl(2,\R)$), therefore it represents an element of $K(\g)$.
An explicit presentation of the algebra $K(\g)$ was found by Berman~\cite{Berman:1989}, whose result we quote in the following theorem (in its version from~\cite{Hainke:2011mt}):
\begin{theorem}
Let $\g$ be a symmetrizable Kac--Moody algebra with generalised Cartan matrix $A$ and let $x_i=e_i-f_i$.
Then the maximal compact subalgebra $K(\g)$ is generated by the elements $\{x_1,\dotsc,x_n\}$ that satisfy the following relations:
\begin{equation}
\label{berman1}
P_{-A_{ij}} (\ad_{x_i})x_j = 0 ,
\end{equation}
where
\begin{equation}
P_m(t) =
\begin{cases}
\bigl(t^2+m^2\bigr) \bigl(t^2+(m-2)^2\bigr) \cdots \bigl(t^2+1\bigr) & \text{if $m$ is odd,} \\
\bigl(t^2+m^2\bigr) \bigl(t^2+(m-2)^2\bigr) \cdots \bigl(t^2+4\bigr) t & \text{if $m$ is even.}
\end{cases}
\end{equation}
\end{theorem}
The theorem drastically simplifies for a simply-laced algebra:
\begin{corollary}
If $\g$ is a simply-laced Kac--Moody algebra, then the generators $\{x_1,\dotsc,x_n\}$ of $K(\g)$ satisfy
\begin{subequations}
\label{berman2}
\begin{align}
[x_i, [x_i,x_j]] + x_j = 0 & \quad \text{if the vertices $i,j$ are connected by a single edge} , \\[2mm]
[x_i,x_j] = 0 & \quad \text{otherwise} .
\end{align}
\end{subequations}
\end{corollary}
To illustrate these general statements we present some examples, which give a 
flavor of how the relations~\eqref{berman1} are concretely realised.
Consider the symmetric algebra $\aethree$, whose Dynkin diagram is in figure~\ref{dynkin:aethree} and whose Cartan matrix is
\begin{equation}
\label{cartan:aethree}
A[\aethree] =
\begin{pmatrix}
2 & -1 & 0 \\
-1 & 2 & -2 \\
0 & -2 & 2 \\
\end{pmatrix}
;
\end{equation}
then its subalgebra $\kaethree$ is generated, thanks to~\eqref{berman1}, by $\{x_1,x_2,x_3\}$ that satisfy
\begin{subequations}
\label{berman:kae3}
\begin{align}
[x_1, [x_1,x_2]] + x_2 = 0 , \quad [x_2, [x_2,x_1]] + x_1 = 0 , \quad [x_1,x_3] = 0 , & \\[1mm]
\label{berman:ex}
[x_2, [x_2, [x_2,x_3]]] + 4 \, [x_2,x_3] = 0 , \quad [x_3, [x_3, [x_3,x_2]]] + 4 \, [x_3,x_2] = 0 . &
\end{align}
\end{subequations}
Let us work out explicitly the first relation in~\eqref{berman:ex}:
since $A_{23}=-2$, from~\eqref{berman1} and the definition of the polynomial $P_m(t)$ we get
\[
\bigl[(\ad_{x_2})^2 + 4\bigr] (\ad_{x_2}) x_3 = (\ad_{x_2})^3 x_3 + 4 (\ad_{x_2}) x_3 \overset{!}{=} 0 ,
\]
which is precisely~\eqref{berman:ex}.
The same construction works for the other relations.

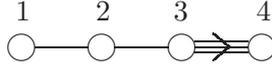
\begin{figure}
\centering
\begin{picture}(60,10)
\put(0,0){\circle{10}}
\put(5,0){\line(1,0){20}}
\put(30,0){\circle{10}}
\put(35,0){\line(1,0){20}}
\put(60,0){\circle{10}}
\put(65,2){\line(1,0){20}}
\put(65,0){\line(1,0){20}}
\put(65,-2){\line(1,0){20}}
\put(90,0){\circle{10}}
\drawline(78,0)(72,5)
\drawline(78,0)(72,-5)
\put(-2,10){$1$}
\put(28,10){$2$}
\put(57,10){$3$}
\put(88,10){$4$}
\end{picture}
\caption{Dynkin diagram of $\gtwo$}
\label{dynkin:gtwo}
\end{figure}

A second example is provided by $\gtwo$, whose Dynkin diagram is given in figure~\ref{dynkin:gtwo}. The corresponding Cartan matrix is
\begin{equation}
\label{cartan:gtwo}
A[\gtwo] =
\begin{pmatrix}
2 & -1 & 0 & 0 \\
-1 & 2 & -1 & 0 \\
0 & -1 & 2 & -1 \\
0 & 0 & -3 & 2 \\
\end{pmatrix}
.
\end{equation}
Therefore the maximal compact subalgebra $\kgtwo$ is generated by $\{x_1,x_2,x_3,x_4\}$ that satisfy
\begin{subequations}
\label{berman:kgtwo}
\begin{align}
[x_1, [x_1,x_2]] + x_2 = 0 , \quad [x_2, [x_2,x_1]] + x_1 = 0 , & \\[1mm]
[x_2, [x_2,x_3]] + x_3 = 0 , \quad [x_3, [x_3,x_2]] + x_2 = 0 , & \\[1mm]
[x_3, [x_3,x_4]] + x_4 = 0 , \quad [x_4, [x_4, [x_4, [x_4,x_3]]] + 10 \, [x_4, [x_4,x_3]] + 9 \, x_3 = 0 , \\[1mm]
[x_1,x_3] = 0 , \quad [x_1,x_4] = 0 , \quad [x_2,x_4] = 0 . &
\end{align}
\end{subequations}
These relations are easily derived from~\eqref{berman1}.

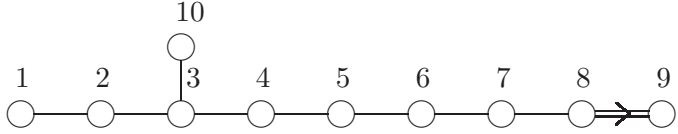
\begin{figure}
\centering
\begin{picture}(300,50)
\put(0,0){\circle{10}}
\put(5,0){\line(1,0){20}}
\put(30,0){\circle{10}}
\put(35,0){\line(1,0){20}}
\put(60,0){\circle{10}}
\put(65,0){\line(1,0){20}}
\put(90,0){\circle{10}}
\put(95,0){\line(1,0){20}}
\put(120,0){\circle{10}}
\put(125,0){\line(1,0){20}}
\put(150,0){\circle{10}}
\put(155,0){\line(1,0){20}}
\put(180,0){\circle{10}}
\put(185,0){\line(1,0){20}}
\put(210,0){\circle{10}}
\put(215,1){\line(1,0){20}}
\put(215,-1){\line(1,0){20}}
\put(240,0){\circle{10}}
\put(60,5){\line(0,1){15}}
\put(60,25){\circle{10}}
\drawline(228,0)(223,4)
\drawline(228,0)(223,-4)
\put(-2,10){$1$}
\put(28,10){$2$}
\put(62,10){$3$}
\put(88,10){$4$}
\put(118,10){$5$}
\put(148,10){$6$}
\put(178,10){$7$}
\put(208,10){$8$}
\put(238,10){$9$}
\put(58,35){$10$}
\end{picture}
\caption{Dynkin diagram of $\be$}
\label{dynkin:be}
\end{figure}

Finally consider the algebra $\be$, whose Dynkin diagram is in figure~\ref{dynkin:be}
(inverting the double arrow, we get the Dynkin diagram of the other rank-10 algebra 
$CE_{10}$, which however does not appear to be relevant for physics).
The simply-laced part of the algebra gives rise to the usual Berman relations~\eqref{berman2} for $\kbe$, and the only one that is not of the type~\eqref{berman2} is
\begin{equation}
\label{berman:be}
[x_9, [x_9, [x_9,x_8]]] + 4 \, [x_9,x_8] = 0 .
\end{equation}

Now we turn back to the general features of the maximal compact subalgebra $K(\g)$.
The subalgebra $K(\g)$ of an infinite-dimensional Kac--Moody algebra 
is \emph{not} of Kac--Moody type~\cite{Kleinschmidt:2005bq}.
This means that there is no grading of $K(\g)$: being the $K(\g)$ generators 
a combination of elements from positive and negative root spaces
\begin{equation}
\label{kroot}
K(\g)_{\alpha} \equiv K(\g)_{-\alpha} \subset \g_\alpha \oplus \g_{-\alpha} ,
\end{equation}
$K(\g)$ possesses a \emph{filtered structure}
\begin{equation}
[K(\g)_{\alpha},K(\g)_{\beta}] \subset K(\g)_{\alpha+\beta} \oplus K(\g)_{\alpha-\beta} .
\end{equation}
This means that we can associate a root to each $K(\g)$ generator, by keeping 
in mind that we are not really speaking of a `root' of $K(\g)$, but of the structure 
given by~\eqref{kroot}. It also means that the established tools of representation 
theory (in particular, highest and lowest weight representations) are not available
for $K(\g)$.

More specifically, below we will look for {\em spinorial}, {\em i.e.}~double-valued 
representations of the compact subalgebra $K(\g)$. For this, what is known 
about  Kac--Moody representation theory is of no help: although it is easy to
generate representations of $K(\g)$ by decomposing representations of $\g$ itself
w.r.t.~$K(\g) \subset \g$,
these will not be spinorial, hence not suitable for the description of
fermions. As it will turn out, so far only non-injective finite-dimensional, hence
\emph{unfaithful}, representations of this type are known, and it remains a 
major open problem to find faithful spinorial representations.

A general property of $K(\g)$ defined through the Cartan--Chevalley involution~\eqref{CCinvolution} is that the restriction of the invariant bilinear form on $\g$ to $K(\g)$ is of \emph{definite} type even though the form on $\g$ is indefinite for Kac--Moody algebras. To understand this, we note that the bilinear form on $\g$ satisfies
\begin{equation}
( e_i , f_j ) = \delta_{ij} ,
\end{equation}
which implies
\begin{equation}\label{normx}
( x_i , x_j ) = -\delta_{ij} ,
\end{equation}
for all simple generators of $K(\g)$, so that the form is negative definite on them
(for $K(\g)$ we obviously do not need the remaining relation $(h_i | h_j) = A_{ij}$).
This property can be seen to extend to all of $K(\g)$. This is another reason why $K(\g)$ is referred to as the maximal compact subalgebra, as finite-dimensional compact (semi-simple) algebras are characterised by having a negative definite invariant form. Unlike for
the finite dimensional  case, the term `compact' in the case of affine or indefinite 
Kac--Moody algebras or groups is used to highlight this 
{\em algebraic} property and does not imply a topological notion of compactness for the 
corresponding algebra or group (even though a topology can be defined w.r.t.~the norm induced by (\ref{normx})).

\section{Spin representations}

We now address the question of constructing spin representations of $K(\g)$, \emph{i.e.}~representations of $K(\g)$ that are double-valued when lifted to a group representation. It turns out that a useful prior step is to introduce a convenient metric on the space of roots, called the DeWitt metric, which arises in canonical treatments of diagonal space-time metrics in general relativity~\cite{DeWitt:1967yk,Damour:2002fz}. It is most easily written in the \emph{wall basis} $e^\ind{a}$ on $\h^*$, such that $\ind{a}=1,\ldots,n$ lies in the same range as the index we have used for labelling the simple roots $\alpha_i$. We use the different font $\ind{a}$ to emphasise  its property that the inner product takes the standard form
\begin{align}
e^\ind{a} \cdot e^\ind{b} = G^{\ind{a}\ind{b}}\,,\quad G^{\ind{a}\ind{b}} = \delta^{\ind{a}\ind{b}} - \frac{1}{n-1} .
\end{align}
The DeWitt metric $G^{\ind{a}\ind{b}}$ has Lorentzian signature $(1,n-1)$. A root $\alpha$ is expanded as $\alpha= \sum_{\ind{a}=1}^n \alpha_{\ind{a}} e^\ind{a}$ and in the examples below we will provide explicit lists of coefficients $\alpha_\ind{a}$ for the simple roots $\alpha_i$ such that their inner product gives back the symmetrised Cartan matrix~\eqref{symC} via $\alpha_i\cdot\alpha_j=B_{ij}$. Below we will also use letters $p_\ind{a} \equiv \alpha_\ind{a}$
to denote the root components, such that $\alpha \equiv (p_1,\cdots,p_d)$.

The spinor representations of $K(\g)$ will be constructed on a tensor product space $V\otimes W$ where $W$ carries the `spinorial' part of the representation and $V$ the `tensorial' part. The spaces $V$ we consider here are obtained within the $s$-fold tensor product of $\h^*$. On $(\h^*)^{\otimes s}$ we can introduce a multi-index $\mathcal{A}=(\ind{a}_1,\ldots,\ind{a}_s)$ and a bilinear form $G^{\mathcal{A}\mathcal{B}} = G^{\ind{a}_1\ind{b}_1} \cdots G^{\ind{a}_s\ind{b}_s}$. We shall typically consider only the case when $V$ is a symmetric tensor product, such that all indices in $G^{\ind{a}_1\ind{b}_1} \cdots G^{\ind{a}_s\ind{b}_s}$ have to be symmetrised. 

The space $W$ on the other hand will be taken to be an irreducible real spinor representation of the Euclidean $\mathfrak{so}(n)$ and we label its components by an index $\alpha$ whose range depends on the size of the real spinor. On $W$ there is also a bilinear form $\delta_{\alpha\beta}$.

Elements of $V\otimes W$ will then be denoted by $\phi^\mathcal{A}_\alpha$ and, for $\mathcal{A}$ referring to the $s$-fold symmetric product, the object $\phi^\mathcal{A}_\alpha$ is called a \emph{spinor} of spin $\bigl(s+\frac{1}{2}\bigr)$; as already emphasised in \cite{Kleinschmidt:2013eka} this 
terminology, which is inspired by $d=4$ higher spin theories, should not be taken too literally 
in that it mixes the notion of space-time spin with that on DeWitt space. Consider now the relation
\begin{equation}
\label{clifford}
\Bigl\{\phi^\mathcal{A}_\alpha,\phi^\mathcal{B}_\beta\Bigr\} = G^\mathcal{AB} \delta_{\alpha\beta} ;
\end{equation}
this equation is meant to generalise the standard Dirac bracket of the components of a gravitino field, see for instance~\cite{Damour:2006xu}. The elements $\phi^\mathcal{A}_\alpha$ are in this case thought of as `second quantised' field operators with canonical anti-commutation relations. A particular way of realising this relation is as Clifford generators~\cite{Lautenbacher:2017}.

Because of the Clifford algebra defining relation~\eqref{clifford}, the $V \otimes W$ matrices 
have to be antisymmetric under simultaneous interchange of the two index pairs.
Given the elements $X_{\mathcal{AB}} = X_{\mathcal{BA}} $, 
$Y_{\mathcal{CD}}= - Y_{\mathcal{CD}}$ in $V$, and 
$S^{\alpha\beta} = - S^{\beta\alpha}$, 
$T^{\gamma\delta} = T^{\delta\gamma}$ in $W$, these satisfy
$X_{\mathcal{AB}} \, S^{\alpha\beta} = -X_{\mathcal{BA}} \, S^{\beta\alpha}$ and $Y_{\mathcal{CD}} \, T^{\gamma\delta} = -Y_{\mathcal{DC}} \, T^{\delta\gamma}$.
The generic elements of the Clifford algebra over $V \otimes W$ can thus always be 
written as a sum $(\hat{A} + \hat{B})$, where
\begin{equation}
\label{operator}
\hat{A} = X_{\mathcal{AB}} \, S^{\alpha\beta} \, \phi^\mathcal{A}_\alpha \phi^\mathcal{B}_\beta \;\;, \quad
\hat{B} = Y_{\mathcal{CD}} \, T^{\gamma\delta} \, \phi^\mathcal{C}_\gamma \phi^\mathcal{D}_\delta .
\end{equation}
The commutator of two such elements is conveniently evaluated in terms of
the `master relation'
\begin{equation}
\label{eq:master}
\Bigl[\hat{A},\hat{B}\Bigr] = \Bigl( [X,Y]_{\mathcal{AB}}\, \{S,T\}^{\alpha\beta} + 
\{X,Y\}_{\mathcal{AB}} \, [S,T]^{\alpha\beta} \Bigr) 
\phi^\mathcal{A}_\alpha \phi^\mathcal{B}_\beta ,
\end{equation}
hence the form~\eqref{operator} is preserved.

We use~\eqref{operator} to write an ansatz for a bilinear operator $J(\alpha)$ associated to a root $\alpha$ of $K(\g)$, {\it viz.}
\begin{equation}
\hat{J}(\alpha) \coloneqq X(\alpha)_{\mathcal{AB}}  \; 
\phi^\mathcal{A}_\alpha \Gamma(\alpha)_{\alpha\beta}\phi^\mathcal{B}_\beta .
\end{equation}
If $\alpha$ is real, $\hat{J}(\alpha)$ refers to the essentially unique element in $K(\g)_\alpha$ while for imaginary roots $\alpha$ there are as many components to $\hat{J}(\alpha)$ as the root multiplicity of $\alpha$. 
The matrix $\Gamma(\alpha)$ is built out of the Euclidean Dirac gamma matrices corresponding to the
$\ell =0$ generators of $K(\g)$, potentially augmented by a \emph{time-like} gamma matrix, in such a way that each element $\alpha$ of the root lattice 
is mapped to an element of the associated real Clifford algebra; its dimension depends on 
which space $W$ we are considering, and is related to the irreducible spinor representation 
of $\mathfrak{so}(n)$ of $\mathfrak{so}(1,n)$ underlying $W$. We shall be more explicit in the examples below. 
$X(\alpha)$ is called \emph{polarisation tensor} and is required to elevate the spin-$\frac12$ 
representation generated by $\Gamma(\alpha)$ to higher spin; again, we shall be more explicit below.

We can express the action of the simple generators $x_i \equiv J(\alpha_i)$ of $K(\g)$ on the spinor $\phi^\mathcal{A}_\alpha$:
this is given by
\begin{equation}
(J(\alpha) \phi)^\mathcal{A}_\alpha  \,\equiv \, \big[\hat{J}(\alpha),\phi\big]^\mathcal{A}_\alpha 
=-2 X(\alpha)^\mathcal{A}{}_{\mathcal{B}} \Gamma(\alpha)_\alpha^\beta \, \phi^\mathcal{B}_\beta ,
\end{equation}
or
\begin{equation}
\label{ansatz}
J(\alpha)_\mathcal{AB}{}^{\alpha\beta} \equiv -2 X(\alpha)_\mathcal{AB} \Gamma(\alpha)^{\alpha\beta} .
\end{equation}
When written without a hat, we think of $\hat{J}(\alpha)$ as the classical representation of the generator that acts simply by matrix multiplication on 
the components of $\phi^\mathcal{A}_\alpha$. 
We will use this expression as our ansatz when for searching for spin representations.
The strategy will be to look for suitable $X(\alpha)$ and $\Gamma(\alpha)$ such that $J(\alpha)$ satisfies the Berman relations~\eqref{berman1}.

\subsection{The simply-laced case}

The ansatz~\eqref{ansatz} allows to translate the Berman relations~\eqref{berman2} of a simply-laced algebra into relations for the matrices $X(\alpha)$ and $\Gamma(\alpha)$.

In the simply-laced case, we can associate a gamma matrix $\Gamma(\alpha)$ to each element $\alpha$ of the root lattice using a standard construction~\cite{Goddard:1986bp}. These matrices $\Gamma(\alpha)$ satisfy the relations
\begin{subequations}
\label{gamma-relations}
\begin{align}
\Gamma(\alpha) \Gamma(\beta) & = (-1)^{\alpha\cdot\beta} \Gamma(\beta) \Gamma(\alpha) , \\
{\Gamma(\alpha)}^T & = (-1)^{\frac{1}{2}\alpha\cdot\alpha} \Gamma(\alpha) , \\
{\Gamma(\alpha)}^2 & = (-1)^{\frac{1}{2}\alpha\cdot\alpha} \id .
\end{align}
\end{subequations}
As we will see below, these relations can also be satisfied when the algebra is not 
simply-laced, provided one normalises the inner product such that the 
shortest roots have $\alpha^2=2$. These relations are absolutely crucial for
the construction of higher spin representations. In particular, the relations (\ref{x-berman})
below would not make sense without (\ref{gamma-relations}).

Using the $\Gamma$-matrices we can construct a solution to the Berman relations~\eqref{berman2} in the non-simply-laced case as follows. First, observe that the expressions~\eqref{gamma-relations} imply
\begin{subequations}
\begin{align}
\lb\Gamma(\alpha), \lb\Gamma(\alpha),\Gamma(\beta)\rb\rb &= -4 \Gamma(\beta) \quad\, \text{when $\alpha\cdot\beta=\mp1$} , \\
\lb\Gamma(\alpha),\Gamma(\beta)\rb &= 0 \qquad\qquad \text{when $\alpha\cdot\beta=0$} .
\end{align}
\end{subequations}
Thus, choosing $J(\alpha_i)=\frac12\Gamma(\alpha_i)$ for all simple roots will solve the Berman relations~\eqref{berman2} for the representation $x_i=J(\alpha_i)$. In terms of the ansatz~\eqref{ansatz}, this means $X(\alpha)=-1/4$ for the polarisation and the space $V$ is one-dimensional. For simply-laced algebras $\g$ one has also that all real roots are Weyl conjugate and we can choose 
\begin{equation}\label{JGamma}
J(\alpha)=\frac12 \Gamma(\alpha) ,
\end{equation}
for all real roots $\alpha$. This is the spin-$\frac12$ representation constructed 
in~\cite{deBuyl:2005zy,Damour:2005zs,Damour:2006xu,Hainke:2011mt}. Its spinorial
character follows from the relation
\begin{equation}
\exp \bigl( 2\pi J(\alpha) \bigr) = - \id ,
\end{equation}
which will also hold for higher spin representations.

Now we move on to higher spin: in this case the space $W$ is more than one-dimensional and the polarisation tensor $X(\alpha)$ has a non-trivial structure. Making the ansatz~\eqref{ansatz} for $J(\alpha)$ we can then determine a sufficient condition for $X(\alpha)$ to solve the Berman relations~\eqref{berman2}. Upon using~\eqref{gamma-relations} and~\eqref{eq:master}, we see that 
\begin{subequations}
\label{x-berman}
\begin{align}
\big\{X(\alpha),X(\beta)\big\} & = \frac{1}{2} X(\alpha\pm\beta) \quad\quad\;\; \text{if $\alpha\cdot\beta=\mp1$} , \\
\big[X(\alpha),X(\beta)\big] & = 0 \qquad\qquad\qquad\quad \text{if $\alpha\cdot\beta=0$} ,
\end{align}
\end{subequations}
is a sufficient condition when the roots range over the simple roots (or even all real roots). In this relation the polarisation tensors indices on the polarisation tensors are raised and lowered with $G^{\mathcal{A}\mathcal{B}}$ and its inverse in order to define the matrix products. The relations~\eqref{x-berman} are a sufficient condition for every simply-laced Kac--Moody algebra and they will be the main equations we solve.
We now discuss higher spin representations for some simply-laced  Kac--Moody algebras, namely $\aed$, $\e$ and $\de$.

\subsubsection{$\aed$ and $\kaed$ ($d>3$)}

The indefinite Kac--Moody algebra $\aed$, obtained by double extension of the Cartan type 
$A_{d-2}$, is the conjectured symmetry algebra for Einstein's gravity in $(d+1)$ dimensions. 
For $d>9$, these algebras are no longer hyperbolic, but of general Lorentzian type. This holds in 
particular for $AE_{10}$, which is a subalgebra of $\e$ and is conjectured to govern 
the gravitational sector of 11-dimensional supergravity.

\begin{figure}
\centering
\begin{picture}(60,30)
\put(0,0){\circle{10}}
\put(5,0){\line(1,0){20}}
\put(30,0){\circle{10}}
\put(35,0){\line(1,0){20}}
\put(60,0){\circle{10}}
\dashline{10}(65,0)(105,0)
\put(110,0){\circle{10}}
\put(70,40){\circle{10}}
\drawline(33.5,3.5)(66.5,36.5)
\drawline(73.5,36.5)(106.5,3.5)
\put(-2,10){$1$}
\put(28,10){$2$}
\put(57,10){$3$}
\put(105,10){$d-1$}
\put(80,40){$d$}
\end{picture}
\caption{Dynkin diagram of $\aed$}
\label{dynkin:aed}
\end{figure}

The $\aed$ Dynkin diagram is depicted in figure~\ref{dynkin:aed}. The corresponding Cartan matrix can be realised by simple roots in the wall basis. Because this algebra is associated with pure Einstein
gravity in $(d+1)$ space-time dimensions, the relevant DeWitt metric is
\begin{equation}
G^\ind{ab} = \delta^\ind{ab} - \frac{1}{d-1} ,
\end{equation}
and the simple roots in the wall basis are
\begin{subequations}
\begin{align}
\alpha_1 & = (1,-1,0,\dotsc,0) , \\
\alpha_2 & = (0,1,-1,0,\dotsc,0) , \\
& \vdots \nonumber \\
\alpha_{d-1} & = (0,...,0,1,-1) ,\\
\alpha_d & = (0,0,1,\dotsc,1,2) .
\end{align}
\end{subequations}
The spin-$\frac{1}{2}$ representation of $\kaed$ can be constructed using the real
matrices $\Gamma(\alpha)$ appearing in~\eqref{gamma-relations}. For their explicit
construction we need the {\em real} SO$(1,d)$ Clifford algebra in $(d+1)$-dimensional  spacetime
generated by $\{\Gamma_\mu, \Gamma_\nu\} = 2 \eta_{\mu\nu}$.\footnote{We use the mostly 
plus signature for Minkowski space-time and $\epsilon^{01\ldots d} =+1$ with Lorentz 
indices $\mu,\nu,... \in\{0,1,\ldots,d\}$. The $\Gamma$-matrices are real
$n_d \times n_d$ matrices with $n_3 =4\,,\, n_4=n_5 =8 \,,\, n_6=\cdots = n_9 = 16$ and $n_{10}=32$ (see
{\em e.g.}~\cite{Polchinski:1998rr,deWit:1992psp}; for yet higher $d$ the numbers follow from Bott periodicity
$n_{d+8} = 16 n_d$~\cite{Atiyah:1964zz}).} 

The SO$(1,d)$ Clifford algebra contains the real element
\begin{equation}
\label{gammastar}
\Gamma_* \equiv \Gamma_0 \Gamma_1\cdots \Gamma_d = \frac{1}{(d+1)!} \epsilon^{\mu_0\mu_1 \ldots \mu_d} \Gamma_{\mu_0} \Gamma_{\mu_1} \cdots \Gamma_{\mu_d} ,
\end{equation}
that satisfies
\begin{equation}
(\Gamma_*)^2 = (-1)^{\frac{d(d+1)}2 +1} , \qquad
\left\{ \begin{array}{rc} 
\textrm{$d$ even:} & \lb \Gamma_* , \Gamma_a \rb =0 \\[2mm]
\textrm{$d$ odd:} & \left\{ \Gamma_* , \Gamma_a \right\} =0 
\end{array}\right\} .
\end{equation}
The {\em real} SO$(d)$ Clifford algebra in $d$ dimensions with $\{ \Gamma_a, \Gamma_b \} 
= 2\delta_{ab}$ (see {\em e.g.}~\cite[Section~2]{deWit:1992psp}), which is associated with the
$\so(d) = K(A_{d-1})$ subalgebra of $K(\aed)$, is obviously contained in the above Clifford algebra. 
The need for the {\em Lorentzian} Clifford algebra SO$(1,d)$ is thus not immediately obvious 
from the above Dynkin diagram, but is necessitated by the extra node $\alpha_d$ associated
with the dual graviton, as we will shortly explain. We also note that when $d\equiv 2\mod 4$, the matrix $\Gamma_*$ is proportional to the identity. 

After these prepararations we define the matrices $\Gamma(\alpha)$ for the simple roots of $AE_d$:
\begin{equation}
\label{Gamma1}
\Gamma(\alpha_1) \equiv  \Gamma_{12} \, , \, \ldots , \, \Gamma(\alpha_{d-1}) \equiv \Gamma_{d-1,d} ,
\end{equation}
for the first $(d-1)$ roots, and
\begin{equation}\label{gamma-def}
\Gamma(\alpha_d) \equiv \Gamma_{345\cdots d-1} \Gamma_* = \Gamma_*  \Gamma_{345\cdots d-1}
\,= (-1)^{\frac{d(d-1)}2} \Gamma_{012d} ,
\end{equation}
for the last simple root.
Before motivating this definition from supergravity, we first check the Berman relations with the identification (\ref{JGamma}).  The relations for the  line spanned by the roots $\alpha_1,\ldots,\alpha_{d-1}$ correspond to the usual $\so(d)$ algebra written in terms of gamma matrices and therefore are automatically satisfied. The  non-trivial key relation to check is  therefore the one involving nodes $d-1$ 
and $d$ of diagram~\ref{dynkin:aed}. This leads to
\begin{equation*}
\begin{split}
\lb \Gamma(\alpha_d) , \lb \Gamma(\alpha_d) , \Gamma(\alpha_{d-1}) \rb\rb &= (\Gamma_*)^2 \lb \Gamma_{34\ldots d-1}, \lb\Gamma_{34\ldots d-1}, \Gamma_{d-1, d}\rb\rb \\
& = 2 \,(\Gamma_*)^2 \lb \Gamma_{34\ldots d-1}, \Gamma_{34\ldots d-2, d}\rb \\\
& = 4 \, (\Gamma_*)^2 (-1)^{d(d-3)/2} \Gamma_{d-1,d} \\
& = -4 \, \Gamma(\alpha_d) ,
\end{split}
\end{equation*}
where the important point is that the property of $\Gamma_*$ is always such that this sign works 
out  to be negative -- showing explicitly the necessity of going Lorentzian, since otherwise the
Berman relations would not hold. In summary, we have verified all Berman relations and 
constructed the spin-$\tfrac12$ representation of $K(AE_d)$ for all $d>3$ in terms of the 
associated Lorentzian Clifford algebra.

The above definition of $\Gamma(\alpha)$ can be extended to the whole root lattice by multiplication. For instance, for real roots $\alpha \equiv (p_1,\cdots,p_d)$ this leads to the simple formula
\begin{equation}
\Gamma(\alpha) = (\Gamma_1)^{p_1} \cdots (\Gamma_d)^{p_d} \, 
(\Gamma_*)^{\frac{1}{d-1}\sum_{a=1}^d p_a }\,.
\end{equation}
These matrices satisfy~\eqref{gamma-relations}, which is crucial for reducing the problem of finding higher spin representations to solve for the polarisation tensors.

Let us briefly motivate the definition~\eqref{gamma-def} from supergravity. The group $AE_d$ is conjectured to be related to symmetries of gravity in $d+1$ space-time dimensions and, where it exists, also of the corresponding supergravity theory.
Decomposing the adjoint of $AE_d$ under its 
$A_{d-1} \cong \mathfrak{sl}(d)$ subalgebra reveals the gravitational fields: at level $\ell=0$ one has the adjoint of $\mathfrak{gl}(d)$ corresponding to the graviton and at level $\ell=1$ one finds the dual graviton that is a hook-tensor representation of type $(d-2,1)$ under $\mathfrak{gl}(d)$, meaning that it can be represented by a tensor $h_{a_1\ldots a_{d-2}, a_0}$ with anti-symmetry in the first $d-2$ indices ($h_{[a_1\ldots a_{d-2}], a_0} = h_{a_1\ldots a_{d-2},a_0}$) and Young irreducibility constraint $h_{[a_1\ldots a_{d-2}, a_0]}=0$.
From the partial match of the bosonic gravitational dynamics to an $AE_d/K(AE_d)$ geodesic one knows this field to be related to (the traceless part of) the spin connection via a duality equation of the form~\cite{Damour:2002cu,Damour:2002et,Kleinschmidt:2005bq}\footnote{More precisely, one should consider the coefficients of anholonomy but for our discussion here this difference does not matter.}
\begin{equation}
\omega_{a_0 bc} = \frac{1}{(d-2)!} \epsilon_{0 bc a_1\ldots a_{d-2} } \, \partial_0 h_{a_1\ldots a_{d-2}, a_0}  .
\end{equation}
The indices $a,b$ here are spatial vector indices of $\so(d)$. The root vector for $\alpha_d$ corresponds to the component $h_{34\ldots d,d}$.

Assuming now a supersymmetry variation for a vector-spinor $\delta_\epsilon \Psi_\mu  = \left(\partial_\mu + \frac14 \omega_{\mu \rho\sigma} \Gamma^{\rho\sigma} \right)\epsilon$ for the standard redefined 
0-component of the vector-spinor leads to~\cite{Damour:2006xu}
\begin{equation}
\delta_\epsilon \left( \Psi_0 - \Gamma_0 \Gamma^a \Psi\right) = \frac14 \omega_{0 ab} \Gamma^{ab}\epsilon - \frac14 \Gamma_0\Gamma^a \omega_{a bc} \Gamma^{bc}\epsilon + \cdots ,
\end{equation}
where the ellipsis denotes terms that contain partial derivative and components of the spin connection that are irrelevant for the present discussion. Using the `dictionary', according to which
$\omega_{0ab}$ multiplies the level $\ell=0$ generator of $K(AE_d)$ acting on the spin-$\tfrac12$ field $\epsilon$, while $\partial_0 h_{a_1\ldots a_{d-2},a_0}$ multiplies the level $\ell=1$ generators of $K(AE_d)$ acting on $\epsilon$ we deduce that the first $d-1$ roots of $K(AE_d)$ should be represented by matrices $\Gamma_{ab}$ as we claimed in~\eqref{Gamma1} and that the generator for the root $\alpha_d$ is represented by
\begin{equation}
\epsilon^{0123\ldots d} \, \Gamma_{012d} \;\propto\;  \Gamma_{345\ldots d-1} \Gamma_* ,
\end{equation}
as claimed in~\eqref{gamma-def}, where we also used the fact that we only consider the traceless part of the spin connection. We recall that when $d\equiv 2 \mod 4$, the matrix $\Gamma_*$ is proportional to the identity and can then be eliminated from this formula. This happens for example for $AE_{10}\subset K(E_{10})$ consistent with the known formulas for the spin-$\tfrac12$ field of $K(E_{10})$, see~\cite{Damour:2006xu} and the next section.

In order to find higher-spin representations of $\kaed$ we can thus look for solutions of~\eqref{x-berman};
the ansatz for $X(\alpha)$ depends on the number of indices $s$ that determine the spin $\bigl(s+\frac{1}{2}\bigr)$.
For the spin-$\frac{3}{2}$ case we have a two-index tensor $X(\alpha)_\ind{ab}$;
the unique solution to~\eqref{x-berman} consistent with the symmetries is
\begin{equation}
X(\alpha)_\ind{a}{}^\ind{b} = -\frac{1}{2} \alpha_\ind{a} \alpha^\ind{b} + \frac{1}{4} \delta_\ind{a}{}^\ind{b} .
\end{equation}
The spin-$\frac{5}{2}$ solution is
\begin{equation}
X(\alpha)_\ind{ab}{}^\ind{cd} = \frac{1}{2} \alpha_\ind{a} \alpha_\ind{b} \alpha^\ind{c} \alpha^\ind{d} - \alpha_\ind{(a} \delta_\ind{b)}{}^\ind{(c} \alpha^\ind{d)} + \frac{1}{4} \delta_\ind{a}^\ind{(c}\delta^\ind{d)}_\ind{b} .
\end{equation}
The spin-$\frac52$ representation is not irreducible: the space of elements of the form $G_{\ind{a}\ind{b}}\phi^{\ind{a}\ind{b}}_\alpha$ forms a subrepresentation isomorphic to the spin-$\frac12$ representation and splits off as a direct summand.
It is worth noting that these expressions do \emph{not} depend on the rank $d$ of the algebra: the coefficients are universal.

In the spin-$\frac{7}{2}$ representation we find a different situation, since now there is an explicit dependence on the rank $d$ (and hence on the dimension of the associated physical theory):
\begin{equation}
\label{spin72:kaed}
\begin{split}
X(\alpha)_\ind{abc}^\ind{\quad def} & = -\frac{1}{3} \alpha_\ind{a} \alpha_\ind{b} \alpha_\ind{c} \alpha^\ind{d} \alpha^\ind{e} \alpha^\ind{f}
+ \frac{3}{2} \alpha_\ind{(a} \alpha_\ind{b} \delta_\ind{c)}^\ind{(d} \alpha^\ind{e} \alpha^\ind{f)}
- \frac{3}{2} \alpha_\ind{(a} \delta_\ind{b}^\ind{(d} \delta_\ind{c)}^\ind{e} \alpha^\ind{f)}
+ \frac{1}{4} \delta_\ind{(a}^\ind{(d} \delta_\ind{b}^\ind{e} \delta_\ind{c)}^\ind{f)} \\
& + \frac{14 + d \pm 2 \sqrt{6(8+d)}}{(2+d)^2} \, \alpha_\ind{(a} G_\ind{bc)} G^\ind{(de} \alpha^\ind{f)} \\
& - \frac{6 \pm \sqrt{6(8+d)}}{6(2+d)} \, \Bigl(\alpha_\ind{(a}\alpha_\ind{b}\alpha_\ind{c)} G^\ind{(de}\alpha^\ind{f)} + \alpha_\ind{(a}G_\ind{bc)} \alpha^\ind{(d}\alpha^\ind{e}\alpha^\ind{f)}\Bigr) .
\end{split}
\end{equation}
Let us also mention that the above formulas are valid for real $\alpha$, and not just the
simple roots, again by Weyl conjugacy (since the Weyl group also acts appropriately
on the polarisation tensors). One can now try to extend these results to still higher spins 
by  making the most general ansatz for a spin-$\frac{9}{2}$ representation in the four-fold symmetric tensor product of $\h^*$. However, we have not been able so far to find
any non-trivial solution to~\eqref{x-berman}, and the same holds for yet
higher spins $s=\frac{11}{2},\frac{13}{2},\cdots$. This clearly indicates the need for more general ans\"atze.

\subsubsection{$\e$ and $\ke$}

The maximally extended hyperbolic Kac--Moody algebra $\e$ is conjectured to be a fundamental symmetry of M-theory, see~\cite{Damour:2002cu}.\footnote{An alternative
  proposal based on the non-hyperbolic `very extended' Kac--Moody algebra $E_{11}$
  was put forward in~\cite{West:2001as}.} Besides its intimate relation with
maximal supergravity, it is also the universal such algebra which contains all other
simply-laced hyperbolic Kac--Moody algebras~\cite{Viswanath}.

\begin{figure}
\centering
\begin{picture}(200,50)
\put(0,0){\circle{10}}
\put(5,0){\line(1,0){20}}
\put(30,0){\circle{10}}
\put(35,0){\line(1,0){20}}
\put(60,0){\circle{10}}
\put(65,0){\line(1,0){20}}
\put(90,0){\circle{10}}
\put(95,0){\line(1,0){20}}
\put(120,0){\circle{10}}
\put(125,0){\line(1,0){20}}
\put(150,0){\circle{10}}
\put(155,0){\line(1,0){20}}
\put(180,0){\circle{10}}
\put(185,0){\line(1,0){20}}
\put(210,0){\circle{10}}
\put(215,0){\line(1,0){20}}
\put(240,0){\circle{10}}
\put(180,5){\line(0,1){15}}
\put(180,25){\circle{10}}
\put(-2,10){$1$}
\put(28,10){$2$}
\put(58,10){$3$}
\put(88,10){$4$}
\put(118,10){$5$}
\put(148,10){$6$}
\put(182,10){$7$}
\put(208,10){$8$}
\put(238,10){$9$}
\put(180,33){$10$}
\end{picture}
\caption{Dynkin diagram of $\e$}
\label{dynkin:e}
\end{figure}
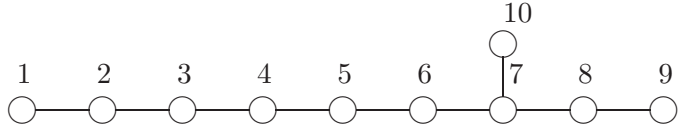

The $\e$ Dynkin diagram is shown in figure~\ref{dynkin:e}.
The DeWitt metric is
\begin{equation}
G^\ind{ab} = \delta^\ind{ab} - \frac{1}{9} ,
\end{equation}
and the simple roots in the wall basis are
\begin{subequations}
\begin{align}
\alpha_1 & = (1,-1,0,\dotsc,0) , \\
\alpha_2 & = (0,1,-1,0,\dotsc,0) , \\
& \vdots \nonumber \\
\alpha_9 & = (0,\dotsc,0,0,1,-1) , \\
\alpha_{10} & = (0,\dotsc,0,1,1,1) .
\end{align}
\end{subequations}
We use the real $32\times32$ gamma matrices that satisfy $\{\Gamma_a,\Gamma_b\}
=2\delta_{ab}$ ($a,b =1,\dotsc,10$) to define
\begin{equation}
\Gamma(\alpha) \coloneqq (\Gamma_1)^{p_1} \cdots (\Gamma_{10})^{p_{10}} .
\end{equation}
or equivalently,
\begin{equation}
\Gamma(\alpha_1) = \Gamma_{12} , \, \dotsc , \, \Gamma(\alpha_9) = \Gamma_{9\,10} , \,
\Gamma(\alpha_{10}) = \Gamma_{89\,10} .
\end{equation}
Then it is easy to check that the generators $J(\alpha)=\frac{1}{2}\Gamma(\alpha)$ 
satisfy the Berman relations~\eqref{berman2}. 

The polarisation tensors of the spin $\frac{3}{2}$, $\frac{5}{2}$ and $\frac{7}{2}$ representations for $\ke$ are exactly the same as in the $\kaed$ case for $d=10$;
in particular, the spin-$\frac{7}{2}$ polarisation tensor is
\begin{equation}
\begin{split}
X(\alpha)_\ind{abc}^\ind{\quad def} & = -\frac{1}{3} \alpha_\ind{a} \alpha_\ind{b} \alpha_\ind{c} \alpha^\ind{d} \alpha^\ind{e} \alpha^\ind{f}
+ \frac{3}{2} \alpha_\ind{(a} \alpha_\ind{b} \delta_\ind{c)}^\ind{(d} \alpha^\ind{e} \alpha^\ind{f)}
- \frac{3}{2} \alpha_\ind{(a} \delta_\ind{b}^\ind{(d} \delta_\ind{c)}^\ind{e} \alpha^\ind{f)}
+ \frac{1}{4} \delta_\ind{(a}^\ind{(d} \delta_\ind{b}^\ind{e} \delta_\ind{c)}^\ind{f)} \\
& + \frac{1}{12}(2\pm\sqrt{3}) \, \alpha_\ind{(a} G_\ind{bc)} G^\ind{(de} \alpha^\ind{f)} \\
& - \frac{1}{12}(1\pm\sqrt{3}) \, \Bigl(\alpha_\ind{(a}\alpha_\ind{b}\alpha_\ind{c)} G^\ind{(de}\alpha^\ind{f)} + \alpha_\ind{(a}G_\ind{bc)} \alpha^\ind{(d}\alpha^\ind{e}\alpha^\ind{f)}\Bigr) .
\end{split}
\end{equation}
These representations have been studied in~\cite{Kleinschmidt:2013eka} and~\cite{Kleinschmidt:2016ivr}.

\subsubsection{$\de$ and $\kde$}
\label{section:de}

We next analyze the spinor representations of $\kde$, the involutory subalgebra of $\de$ which is related 
to the pure type I supergravity in 10 dimensions~\cite{Hillmann:2006ic}.

\begin{figure}
\centering
\begin{picture}(200,50)
\put(0,0){\circle{10}}
\put(5,0){\line(1,0){20}}
\put(30,0){\circle{10}}
\put(35,0){\line(1,0){20}}
\put(60,0){\circle{10}}
\put(65,0){\line(1,0){20}}
\put(90,0){\circle{10}}
\put(95,0){\line(1,0){20}}
\put(120,0){\circle{10}}
\put(125,0){\line(1,0){20}}
\put(150,0){\circle{10}}
\put(155,0){\line(1,0){20}}
\put(180,0){\circle{10}}
\put(185,0){\line(1,0){20}}
\put(210,0){\circle{10}}
\put(180,5){\line(0,1){15}}
\put(180,25){\circle{10}}
\put(60,5){\line(0,1){15}}
\put(60,25){\circle{10}}
\put(-2,10){$1$}
\put(28,10){$2$}
\put(62,10){$3$}
\put(88,10){$4$}
\put(118,10){$5$}
\put(148,10){$6$}
\put(182,10){$7$}
\put(208,10){$8$}
\put(180,33){$9$}
\put(60,33){$10$}
\end{picture}
\caption{Dynkin diagram of $\de$}
\label{dynkin:de}
\end{figure}
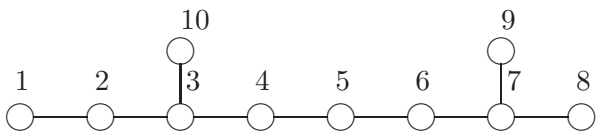

The $\de$ Dynkin diagram is given in figure~\ref{dynkin:de}, and its DeWitt metric is
\begin{equation}
G^\ind{ab} = \delta^\ind{ab} - \frac{1}{8} , \quad G^{10,\ind{a}} = G^{\ind{a},10} = 0 , \quad G^{10,10} = 2 ,
\end{equation}
where $\ind{a,b}=1,\dotsc,9$; in this case the metric takes a non-standard form because 
of the presence of the dilaton in the supergravity theory associated to $G^{10,10}$.

The ten simple roots are
\begin{subequations}
\begin{align}
\alpha_1 & = (1,-1,0,\dotsc|0) , \\
\alpha_2 & = (0,1,-1,0,\dotsc|0) , \\
& \vdots \nonumber \\
\alpha_8 & = (0,\dotsc,0,1,-1|0) , \\
\alpha_9 & = (0,\dotsc,0,1,1|1/2) , \\
\alpha_{10} & = (0,0,0,1,\dotsc,1\,| -1/2) .
\end{align}
\end{subequations}
To find the spin-$\frac{1}{2}$ representation we make use of the $16\times16$ real SO(9) 
gamma matrices $\Gamma_a$ ($a =1,\dotsc,9$) and define 
(for $\alpha \equiv  (p_1,\ldots, p_9|p_{10})$)
\begin{equation}
\Gamma(\alpha) \coloneqq
(\Gamma_1)^{p_1} \cdots (\Gamma_9)^{p_{9}} ,
\end{equation}
so that, in particular, $\Gamma(\alpha_{10}) = \Gamma_{456789}$.  Importantly, the dilaton coordinate (the tenth one) does not 
affect the representation. It is easily checked that $J(\alpha)=\frac{1}{2}\Gamma(\alpha)$ 
satisfies the Berman relations for a simply laced algebra~\eqref{berman2}. The polarisation 
tensors for the higher spin representations are the same of the $\e$ case, $\de$ 
being a simply-laced algebra.

We note that the above definition implies $\Gamma(\alpha_8) = \Gamma(\alpha_9)=\Gamma_{89}$.
This degeneracy follows from the fact that $\de$ is associated with type I (half maximal)
supergravity, and there are only $16$, rather than $32$, real spinor components.
It means that this $16$-dimensional spinor representation of $K(\de)$ does not distinguish the two spinor nodes of the Dynkin diagram. From 
the point of view of space-time rotations this is related to the fact that the spatial rotations SO(9) 
sit inside SO(9)$\times$ SO(9) $\subset K(\de)$ as the \emph{diagonal} subgroup, and 
thus treat both nodes in the same way. One could also introduce a bit artificially a $16$-component spinor representation where $\Gamma(\alpha_9)=-\Gamma(\alpha_8)$ that is superficially of different chirality, but this change of sign can be undone by simply redefining the generators of $K(\de)$ and thus is of real significance.
As usual assigning chirality is a convention that only becomes meaningful in relation to other chiral objects. 

\subsection{The non-simply-laced case}

When the algebra $\g$ is not simply-laced we have to solve different relations than~\eqref{berman2}. This is, we look for generators $J(\alpha)$ that satisfy the original Berman relations~\eqref{berman1}.

\subsubsection{$\aethree$ and $\kaethree$}

The Dynkin diagram of $\aethree$ is given in figure~\ref{dynkin:aethree} and its Cartan matrix in~\eqref{cartan:aethree}. It was first studied in detail in~\cite{FeingoldFrenkel}. 
We notice that this algebra is non-simply-laced but symmetric:
this means that the Berman relations are symmetric, and they are given in~\eqref{berman:kae3}.

The metric is
\begin{equation}
G^\ind{ab} = \delta^\ind{ab} - \frac{1}{2} ,
\end{equation}
and the simple roots in the wall basis are
\begin{subequations}
\begin{align}
\alpha_1 & = (1,-1,0) , \\
\alpha_2 & = (0,1,-1) , \\
\alpha_3 & = (0,0,2) .
\end{align}
\end{subequations}
We construct the gamma matrices $\Gamma(\alpha)$ starting from the real symmetric $4\times4$ gamma matrices $\Gamma_a$ ($a =1,\dotsc,3$) that satisfy $\{\Gamma_a,\Gamma_b\}=
2\delta_{ab}$.
For any root $\alpha \equiv (p_1,p_2,p_3)$ we deduce the matrices $\Gamma(\alpha)$ as
a special case of~\eqref{gamma-def}:
\begin{equation}
\Gamma(\alpha) = (\Gamma_1)^{p_1} (\Gamma_2)^{p_2} (\Gamma_3)^{p_3} (\Gamma_*)^{(p_1+p_2+p_3)/2} .
\end{equation}
This definition is chosen in such a way that the relations~\eqref{gamma-relations} 
are satisfied for any root $\alpha$, as can be verified.
Then, the matrices for the simple roots are $\Gamma(\alpha_1)=\Gamma_{12}$, $\Gamma(\alpha_2)=\Gamma_{23}$ and $\Gamma(\alpha_3)=\Gamma_*$. It is easy
to see that these matrices generate the algebra $\mathfrak{u}(2)$.

Now it is easy to check that the generators $J(\alpha)=\frac{1}{2}\Gamma(\alpha)$ give rise to a spin-$\frac{1}{2}$ representation as in the simply-laced case, since they satisfy the Berman relations~\eqref{berman:kae3}. Because $\Gamma_*$ commutes  with all elements generated 
from the simple roots, there exists in  this particular case a  non-trivial {\em central element} 
in the second quantised description which is  given by 
\begin{equation}
{\cal Z} \,=\, \phi^\ind{a}_\alpha G_{\ind{ab}} (\Gamma_*)_{\alpha\beta} \phi^\ind{b}_\beta .
\end{equation}
For higher $d$ such a central element exists whenever $\Gamma_*$ is anti-symmetric
(because the fermion operators {\em anti}-commute), which
for $AE_d$ is the case for $d=3,7,\cdots$. However, it does not exist for $\ke$. The existence of 
this central element plays an important role in the analysis of supersymmetric quantum
cosmology in a Bianchi-type truncation \cite{Damour:2013eua,Damour:2014cba}, where
it helps in particular in diagonalising the Hamiltonian and in elucidating the structure
of the quartic fermion terms (which are often ignored in discussions of supersymmetric
quantum cosmology).

In order to find higher spin representations, we need to rewrite the Berman relations~\eqref{berman:kae3} in terms of the polarisation tensors $X(\alpha)$:
from the ansatz~\eqref{ansatz} and the relations~\eqref{gamma-relations} among the $\Gamma(\alpha)$'s, the equations~\eqref{berman:kae3} turn into
\begin{subequations}
\begin{align}
4 \{X(\alpha_1), \{X(\alpha_1),X(\alpha_2)\}\} - X(\alpha_2) & = 0 , \\
4 \{X(\alpha_2), \{X(\alpha_2),X(\alpha_1)\}\} - X(\alpha_1) & = 0 , \\
\{X(\alpha_2), \{X(\alpha_2), \{X(\alpha_2),X(\alpha_3)\}\} - \{X(\alpha_2),X(\alpha_3)\} & = 0 , \\
\{X(\alpha_3), \{X(\alpha_3), \{X(\alpha_3),X(\alpha_2)\}\} - \{X(\alpha_3),X(\alpha_2)\} & = 0 , \\
[X(\alpha_1),X(\alpha_3)] & = 0 .
\end{align}
\end{subequations}
There are several solutions to these relations.
The spin-$\frac{3}{2}$ and spin-$\frac{5}{2}$ representations are exactly the same as in the simply-laced case.
The novelty of $\kaethree$ stands in the spin-$\frac{7}{2}$, which gives \emph{two} different representations:
\begin{equation}
\begin{split}
X(\alpha)_\ind{abc}^\ind{\quad def} & = -\frac{1}{3} \alpha_\ind{a} \alpha_\ind{b} \alpha_\ind{c} \alpha^\ind{d} \alpha^\ind{e} \alpha^\ind{f}
+ \frac{3}{2} \alpha_\ind{(a} \alpha_\ind{b} \delta_\ind{c)}^\ind{(d} \alpha^\ind{e} \alpha^\ind{f)}
- \frac{3}{2} \alpha_\ind{(a} \delta_\ind{b}^\ind{(d} \delta_\ind{c)}^\ind{e} \alpha^\ind{f)}
+ \frac{1}{4} \delta_\ind{(a}^\ind{(d} \delta_\ind{b}^\ind{e} \delta_\ind{c)}^\ind{f)} \\
& + \frac{1}{25}(17\pm2\sqrt{66}) \, \alpha_\ind{(a} G_\ind{bc)} G^\ind{(de} \alpha^\ind{f)} \\
& - \frac{1}{30}(6\pm\sqrt{66}) \, \Bigl(\alpha_\ind{(a}\alpha_\ind{b}\alpha_\ind{c)} G^\ind{(de}\alpha^\ind{f)} + \alpha_\ind{(a}G_\ind{bc)} \alpha^\ind{(d}\alpha^\ind{e}\alpha^\ind{f)}\Bigr) ,
\end{split}
\end{equation}
and
\begin{equation}
\begin{split}
X(\alpha)_\ind{abc}^\ind{\quad def} & = -\frac{1}{3} \alpha_\ind{a} \alpha_\ind{b} \alpha_\ind{c} \alpha^\ind{d} \alpha^\ind{e} \alpha^\ind{f}
+ \frac{3}{2} \alpha_\ind{(a} \alpha_\ind{b} \delta_\ind{c)}^\ind{(d} \alpha^\ind{e} \alpha^\ind{f)}
- \frac{3}{2} \alpha_\ind{(a} \delta_\ind{b}^\ind{(d} \delta_\ind{c)}^\ind{e} \alpha^\ind{f)}
+ \frac{1}{4} \delta_\ind{(a}^\ind{(d} \delta_\ind{b}^\ind{e} \delta_\ind{c)}^\ind{f)} \\
& + \frac{1}{50}(19\pm4\sqrt{21}) \, \alpha_\ind{(a} G_\ind{bc)} G^\ind{(de} \alpha^\ind{f)} \\
& - \frac{1}{30}(6\pm\sqrt{21}) \, \Bigl(\alpha_\ind{(a}\alpha_\ind{b}\alpha_\ind{c)} G^\ind{(de}\alpha^\ind{f)} + \alpha_\ind{(a}G_\ind{bc)} \alpha^\ind{(d}\alpha^\ind{e}\alpha^\ind{f)}\Bigr) .
\end{split}
\end{equation}
Note that the first solution can be obtained from the $\kaed$ case by setting $d=3$ (compare with equation~\eqref{spin72:kaed}), while the second expression is completely new. Its relation to gravity is unknown to date.

\subsubsection{$\gtwo$ and $\kgtwo$}

$\gtwo$ plays the role of the symmetry algebra for 5-dimensional supergravity, similarly as 
$\e$ does for 11-dimensional supergravity~\cite{Mizoguchi:2005zf}.
There are indeed many similarities between the two theories~\cite{Mizoguchi:1998wv};
nevertheless, the underlying symmetry algebras $\gtwo$ and $\e$ exhibit very different properties.

The $\gtwo$ Dynkin diagram is in figure~\ref{dynkin:gtwo}, and its Cartan matrix in~\eqref{cartan:gtwo}.
This algebra is not only non-simply-laced, it is also non-symmetric.
Nonetheless $\gtwo$ is symmetrizable, which means that it is possible to find a diagonal matrix $D$ such that $B=DA$ is symmetric.
The choice $D=\diag\bigl(3,3,3,1\bigr)$ leads to
\begin{equation}
B =
\begin{pmatrix}
6 & -3 & 0 & 0 \\
-3 & 6 & -3 & 0 \\
0 & -3 & 6 & -3 \\
0 & 0 & -3 & 2
\end{pmatrix}
;
\end{equation}
since the scalar product will be introduced with respect to the symmetrised Cartan matrix $B$, the fourth root $\alpha_4$ will be interpreted as the `short' root with length $2$.
The Berman relations are given in~\eqref{berman:kgtwo}.

Because the associated supergravity theory lives in five 
dimensions~\cite{Chamseddine:1980sp,Cremmer:1980gs}, the DeWitt metric is
\begin{equation}
\label{dewitt:g2}
G^\ind{ab} = \,\delta^\ind{ab} - \frac13 .
\end{equation}
However, in order to ensure that the root lengths are always even and not fractional,
we rescale this metric by a factor of 3, {\it viz.}
\begin{equation}
\label{dewitt:g2new}
G^\ind{ab} \; \rightarrow \tilde{G}^\ind{ab} \equiv 3 G^\ind{ab}   = 3 \,\delta^\ind{ab} - 1 .
\end{equation}
A further advantage of this choice (which of course does not affect the Cartan matrix of $G_2^{++}$)
is that relations (\ref{gamma-relations}) remain in force. 

The simple roots in the wall basis are given by
\begin{subequations}
\begin{align}
\alpha_1 & = (1,-1,0,0) , \\
\alpha_2 & = (0,1,-1,0) , \\
\alpha_3 & = (0,0,1,-1) , \\
\alpha_4 & = (0,0,0,1) .
\end{align}
\end{subequations}
For the construction of the associated $\Gamma(\alpha_i)$ we use the very same 
8-by-8 matrices as for $AE_4$, the only difference being the assignment for the 
simple root $\alpha_4$:
\begin{equation}\label{GammaG2}
\Gamma(\alpha_1) = \Gamma_{12} , \; \Gamma(\alpha_2) = \Gamma_{23} , \;
\Gamma(\alpha_3) = \Gamma_{34} , \; \Gamma(\alpha_4) = \Gamma_4\Gamma_* ,
\end{equation}
where $\Gamma_*$ was defined in~\eqref{gammastar} and commutes with $\Gamma_a$.  As before its presence is necessary to satisfy the quadrilinear relation in (\ref{berman:kgtwo}).
We recall that the R symmetry of simple supergravity in five dimensions 
is USp(2) ~\cite{Cremmer:1980gs}. As we shall see, this R symmetry will be enlarged
to USp(4) by the extension to $K(G_2^{++})$. We also note that a proposal for including 
the USp(2) R symmetry into the algebra by using a non-split real form of an 
extension of $G_2^{++}$ was made in~\cite{Kleinschmidt:2008jj}.

In order to investigate the higher spin representations, we write the Berman relations for the polarisation tensors $X(\alpha)$:
they are
\begin{subequations}
\begin{align}
4 \{X(\alpha_1), \{X(\alpha_1),X(\alpha_2)\}\} - X(\alpha_2) & = 0 , \\
4 \{X(\alpha_2), \{X(\alpha_2),X(\alpha_1)\}\} - X(\alpha_1) & = 0 , \\
4 \{X(\alpha_2), \{X(\alpha_2),X(\alpha_3)\}\} - X(\alpha_3) & = 0 , \\
4 \{X(\alpha_3), \{X(\alpha_3),X(\alpha_2)\}\} - X(\alpha_2) & = 0 , \\
4 \{X(\alpha_3), \{X(\alpha_3),X(\alpha_4)\}\} - X(\alpha_4) & = 0 , \\
\begin{split}
16 \{X(\alpha_4), \{X(\alpha_4), \{X(\alpha_4), \{X(\alpha_4),X(\alpha_3)\}\}\}\} & \\
- 40 \{X(\alpha_4), \{X(\alpha_4),X(\alpha_3)\}\} + 9 X(\alpha_3) & = 0 ,
\end{split} \\
[X(\alpha_1),X(\alpha_3)] = [X(\alpha_1),X(\alpha_4)] = [X(\alpha_2),X(\alpha_4)] & = 0 .
\end{align}
\end{subequations}
Up to now, we have considered ans\"atze for $X(\alpha)$ in which the coefficients are the same for all the simple roots;
in this case, because of the short root $\alpha_4$, it is natural to assume a different behaviour according to the length of the roots.
This means that we expect to find different coefficients for $X(\alpha_4)$ if compared to the other polarisation tensors.
The spin-$\frac{3}{2}$ representation is indeed found to be
\begin{subequations}
\begin{align}
X(\alpha)_\ind{a}{}^\ind{b} & = \mp \frac{1}{2} \alpha_\ind{a} \alpha^\ind{b} \pm \frac{1}{4} \delta_\ind{a}{}^\ind{b} \quad \text{if $\alpha\ne\alpha_4$} , \\
X(\alpha)_\ind{c}{}^\ind{d} & = \pm \frac{3}{2} \alpha_\ind{c} \alpha^\ind{d} \mp \frac{3}{4} \delta_\ind{c}{}^\ind{d} \quad \text{if $\alpha=\alpha_4$} .
\end{align}
\end{subequations}
It is important to stress that the first solution is consistent with the results from the simply-laced case:
indeed $\alpha_1,\alpha_2,\alpha_3$ form a simply-laced part of $\kgtwo$.

\subsubsection{$\be$ and $\kbe$}

As a last example we consider the algebra $\kbe$, which is related to the low energy limit 
of the heterotic and type I superstrings~\cite{Damour:2000hv}.
One can here proceed in the usual way, by looking for gamma matrices that satisfy 
the Berman relations and then look for the polarisation tensors.
It is indeed possible to find at least one matrix representation that fullfills the 
Berman relations, which however suffers from the same degeneracy as $\de$:
the contribution of the extended root $\alpha_9$ (refer to the Dynkin diagram~\ref{dynkin:be}) has to be trivially realised if one wants to obtain a representation consistent with the physical theory.
This unusual feature, however, does not lead to inconsistencies or to a trivialization of the full algebra:
the Berman relation~\eqref{berman:be} makes possible to have $x_9=0$ without affecting the other generators.
This is a peculiar property of this particular relation, since $x_9$ appears in both commutators in~\eqref{berman:be}.

Once one has fixed the generator associated to $\alpha_9$ to be zero, then the spin representation is easily found:
``removing'' the ninth node we are left with the affine algebra $E_9$, and the $K(E_9)$ representations are known~\cite{Nicolai:2004nv,Kleinschmidt:2006dy}.
They are given by the usual ansatz for $\Gamma(\alpha)$ (the same as for $\de$), built out of the $16\times16$ Dirac matrices.
The polarisation tensors are then the standard ones for a simply-laced algebra.

Finally, let us comment on the consistency of the representation by focusing 
on a particular subalgebra of $\be$, namely $\de$.
The embedding $\de\subset\be$ is suggested by the embedding of pure type I supergravity
into type I supergravity with additional matter couplings. This embedding can be directly realised by
comparison of the Dynkin diagrams~\ref{dynkin:be} and~\ref{dynkin:de}:
let $E_i$, $H_i$, $F_i$ be the $\de$ generators and $e_i$, $h_i$, $f_i$ the $\be$ ones.
By letting
\begin{subequations}
\label{embedd}
\begin{align}
E_i = e_i, \quad H_i = h_i, \quad F_i = f_i \qquad \text{for } i = 1,\dotsc,8,10 , & \\
E_9 = \frac{1}{2}[e_9,[e_9,e_8]], \quad H_9 = h_8 + h_9, \quad F_9 = \frac{1}{2}[f_9,[f_9,f_8]], &
\end{align}
\end{subequations}
one can check that the $\de$ Chevalley--Serre relations are satisfied by virtue of the $\be$ ones.

By using the explicit form of the $\kbe$ generators $x_i=e_i-f_i$ we find that
\[
[x_9,[x_9,x_8]] = [e_9,[e_9,e_8]] - [f_9,[f_9,f_8]] - 2 x_8 ,
\]
which, by virtue of the embedding~\eqref{embedd}, is equivalent to
\begin{equation}
X_9 = \frac{1}{2} [x_9,[x_9,x_8]] + x_8 ,
\end{equation}
where $X_i=E_i-F_i$ are the $\kde$ generators.
It is clear, from the last expression, that $x_9=0$ implies $X_9=x_8$:
this is consistent with the results from section~\ref{section:de}, where the gamma matrices 
relative to the roots $\alpha_8$ and $\alpha_9$ are equal,
$\Gamma(\alpha_8)=\Gamma(\alpha_9)=\Gamma_{89}$, which means 
that the corresponding generators are equivalent. We here see the same 
degeneracy that we already encountered for $\de$, in accord with the fact
that $\be$ is relevant for type I supergravity with (vector multiplet) matter couplings.

\section{Quotients}

The spin representations constructed above are unfaifthful, since they are finite-dimensional
representations of an infinite-dimensional Lie algebra.
The unfaithfulness implies the existence of non-trivial ideals $\mathfrak{i}$ for the algebra, 
where $\mathfrak{i}$ is spanned by those combinations of $K(\g)$ elements which 
annihilate all vectors in the given representation. For each $\mathfrak{i}$ this entails 
the existence of the quotient Lie algebra
\begin{equation}
\mathfrak{q} = K(\g) / \mathfrak{i} .
\end{equation}
Growing with the spin of the representation the algebra gets more and more faithful, so that the quotient becomes bigger (and the ideal smaller). For instance, in the case of the 
spin-$\frac{1}{2}$ representation of $\ke$ the quotient is isomorphic to $\so(32)$, as can 
be easily shown by using
\begin{equation}
\label{iso}
\mathfrak{q}\,\cong\, \im\rho ,
\end{equation}
where $\rho:K(\g)\to \mathrm{End}(V\otimes W)$ is the representation map. It then
follows immediately that  $\mathfrak{q}\cong \so(32)$ because the matrices 
$\Gamma(\alpha)$ for real roots $\alpha$ close into the antisymmetric 32-by-32 matrices.
Things are not so easy for the higher spin representations, due to the dimension of 
the representations and to the presence of non-trivial polarisation tensors.

In general, questions about the quotient algebra can be asked in terms of the representation matrices because these give a faithful image of $\mathfrak{q}$. However, we will
encounter two very unusual features. One is that the quotient algebra $\q$ in general is 
{\em not} a subalgebra  of $K(\g)$. The other is that, as a finite-dimensional Lie algebra,
$\q$ is in general {\em non-compact} even though it descends from the maximal
compact subalgebra $K(\g)\subset \g\,$!

\subsection{Spin-$\frac12$ quotients in $AE_d$}

From the description of the simple generators for $K(AE_d)$ in the spin-$\tfrac12$ representation in~\eqref{Gamma1} and~\eqref{gamma-def} it is also possible to determine the quotient algebras for the various $d$. The quotient depends on the properties of the gamma matrices of $\so(1,d)$ that are known to exhibit Bott periodicity, \emph{i.e.}, the properties (Majorana, Weyl, etc.) are periodic in $d$ with period $8$.

The relevant quotient algebra in terms of gamma matrices is generated by 
$\Gamma_{ab}$ ($\ell=0$), $\Gamma_{0abc}$ ($\ell=1$) and repeated commutation for the various cases leads to antisymmetric matrices according to
\begin{subequations}
\begin{align}
\ell&=0&:&&  \Gamma_{[2]} \\
\ell&=1&:&& \Gamma_0 \Gamma_{[3]} \\
\ell&=2\subset\lb \ell=1,\ell=1\rb & :&&  \Gamma_{[6]} \\
\ell&=3\subset\lb \ell=2,\ell=1\rb & :&&  \Gamma_0 \Gamma_{[7]} \\
\ell&=4\subset\lb \ell=3,\ell=1\rb & :&&  \Gamma_{[10]} \\
&\quad \vdots &&& \vdots\phantom{xx} \nonumber
\end{align}
\end{subequations}
We have only listed the new matrices at each step and for a given $d$ one has to truncate away those matrices that vanish by having too many antisymmetric indices. $\Gamma_{[k]}$ here denotes the $k$-fold antisymmetric product of gamma matrices. 

Based on this sequence of generators one can determine the quotient Lie algebras as
\begin{subequations}
\begin{align}
\qh (K(AE_3)) &\cong \mathfrak{u}(2)  \qquad\qquad\qquad\quad\;\;\;\, \text{acting on $\R^4$}
\\ 
\qh (K(AE_4)) & \cong   \usp(4) \cong \so(5)  \qquad\quad\;\; \text{acting on $\R^8$}           
\\ 
\qh (K(AE_5)) &\cong   \usp(4) \oplus \usp(4)  \qquad\;\;\;\; \text{acting on $\R^8\oplus \R^8$}           
\\ 
\qh (K(AE_6)) &\cong  \mathfrak{spin}(9)  \qquad\qquad\qquad\;\;\; \text{acting on $\R^{16}$}           
\\ 
\qh (K(AE_7)) &\cong  \mathfrak{u}(8)  \qquad\qquad\qquad\quad\;\;\;\; \text{acting on $\R^{16}$}           
\\ 
\qh (K(AE_8)) &\cong   \so(16)  \qquad\qquad\qquad\quad\, \text{acting on $\R^{16}$}           
\\ 
\qh (K(AE_9)) &\cong   \so(16) \oplus \so(16) \qquad\quad\; \text{acting on $\R^{16}\oplus \R^{16}$}           
\\ 
\qh (K(AE_{10})) &\cong   \so(32)   \qquad\qquad\qquad\quad\, \text{acting on $\R^{32}$}
\end{align}
\end{subequations}
All spinorial representations are double valued as it should be. As we have indicated in the above list, in the cases when the number of space dimensions is $d=5+4k$ (for $k\in\Z_{\geq 0}$), one may naively end up with a reducible spinor space when constructing the Clifford algebra. For these values of $d$ one has that $(\Gamma_*)^2=+\id$ and can use this to define chiral spinors by projecting to the $\Gamma_*=\pm \id$ eigenspaces. The irreducible spin-$\tfrac12$ representation then just corresponds to a single summand of the space given in the list. This is the case in particular for $K(AE_9)$ that embeds into $K(DE_{10})$ where we constructed the $16$-dimensional spin-$\frac12$ representation in section~\ref{section:de}.

The occurrence of unitary algebras for $d=3+4k$ is due to the fact that there the element $\Gamma_*$ belongs to the quotient and since $(\Gamma_*)^2=-\id$ this can be used as a complex structure that is respected by all elements of the quotient, leading to a unitary algebra.

We also note some coincidences between quotient algebras that have a physical interpretation in terms of truncations of certain mulitplets in supergravity:
\begin{subequations}
\begin{align}
\qh (K(G_2^{++})) &\cong \qh(K(AE_4)) \cong \usp(4) , \\
\qh(K(DE_{10})) &\cong \qh(K(AE_{9})) \cong \qh(K(BE_{10})) \cong \so(16) , \\
\qh(K(AE_{10})) &\cong \qh(K(\e)) \cong \so(32) .
\end{align}
\end{subequations}
Finally, we note that for all these spin-$\frac12$ representations {\em only the real root
spaces are represented faithfully}, but not the root spaces associated with imaginary roots.

\subsection{Quotient algebras for $\ke$}

For $\ke$, we present a more detailed analysis of the quotient Lie algebras for the higher spin representations. 
The various representations of $\ke$ that we have presented above admit an invariant bilinear form $(\phi,\psi)$, whose specific form depends on the representation. 
The form is invariant w.r.t.~the action of the algebra elements, or equivalently
\begin{equation}
\label{invariant}
(x\cdot\phi,\psi) + (\phi, x\cdot\psi) = 0 ,
\end{equation}
where $x\cdot$ represents the action of a $\ke$ element on the representation space. The bilinear form is symmetric in its arguments.

The abstract expression~\eqref{invariant} can be written explicitly in terms of the generators of the representation, and the general result is
\begin{equation}
\label{constraint}
\eta J + J^T \eta = 0 ,
\end{equation}
where $\eta \equiv G^\mathcal{AB}\delta_{\alpha\beta}$ is the bilinear form and $J\equiv J(\alpha)$ is the root generator.
Plugging the ansatz~\eqref{ansatz} into~\eqref{constraint}, one finds for an antisymmetric $\Gamma$ (\emph{e.g.}~for a real root)
\begin{subequations}
\label{constsplit}
\begin{equation}
GX - X^T G = 0 ,
\end{equation}
which means \emph{symmetric} $X$; on the converse, for a symmetric $\Gamma$ (\emph{e.g.}~for a null root)
\begin{equation}
GX + X^T G = 0 ,
\end{equation}
\end{subequations}
\emph{i.e.}~\emph{antisymmetric} $X$.

We note that not every involutory subalgebra of an indefinite Kac--Moody algebra will
admit such an invariant form. For example, for $K(E_9)$ and $K(E_{11})$ there are no such invariant bilinear forms~\cite{Damour:2006xu}. We also note that for $E_{11}$ one normally considers an involution different from~\eqref{CCinvolution}, called the `temporal' involution~\cite{West:2001as,Englert:2003py,Keurentjes:2004bv} such that the fixed-point algebra has an indefinite bilinear form and in particular contains a Lorentz subalgebra $\mathfrak{so}(1,10)$ rather than the Euclidean $\mathfrak{so}(11)$.As these satisfy~\eqref{constraint}, we immediately deduce that $\mathfrak{q}\subset \mathfrak{so}(\eta)$, where $\eta$ is the bilinear form on the representation space that is left invariant by the action of $K(\g)$. In the following, we shall analyse this in some detail for the case of $\e$.

In the spin-$\frac{1}{2}$ case, the bilinear form is $\eta=\delta_{\alpha\beta}$ and its signature is $(32,0)$: the quotient algebra must therefore be a subalgebra of $\so(32)$. By inspection of the explicit representation matrices in terms of anti-symmetric gamma matrices one easily checks that any $\so(32)$ matrix (in the fundamental representation) is in the image of the spin-$\frac12$ representation and therefore the quotient algebra is isomorphic to $\so(32)$. At the group level, the corresponding isomorphism is $K(E_{10})= \SO(32)$.\footnote{ For $K(E_{11})$, the corresponding quotient Lie algebra is $\mathfrak{sl}(32)$~\cite{West:2003fc}. }

For the higher spin representations, it is less easy to compute explicitly a basis for the image of the representation (which is isomorphic to the quotient by~\eqref{iso}).
The dimension of the expected quotient is usually big, and it is a very hard computational problem to find all the independent matrices which form a basis for the quotient;
furthermore the dimension of the representation itself is quite large, which 
means a huge size for the matrices.

In order to overcome these problems, we make use of the following procedure:
given the $\Gamma(\alpha)$, the constraint~\eqref{constraint} can be solved for the $X(\alpha)$.
If the solutions $X$ are realised in the algebra $\ke$, then they provide a basis  (together with $\Gamma(\alpha)$) for the image of the representation. Hence, by~\eqref{iso}, 
they provide a basis for the quotient.
From a practical point of view, we have to choose a basis of $\Gamma$ matrices (which is made of 1024 elements), and then solve the constraint~\eqref{constraint} with respect to $X$:
therefore we have to check that all the solutions are realised in the representation space.

\subsubsection{Spin-$\frac{3}{2}$}

The bilinear form is given by
\begin{equation}
(\phi,\psi) = \phi^\ind{a} G_\ind{ab} \psi^\ind{b} ,
\end{equation}
whose signature is $(288,32)$. The dimension of $\so(288,32)$ is $51\,040$, that we can decompose into $J=X\otimes \Gamma$ as follows. 
The $\Gamma(\alpha)$ range over the full Clifford algebra in ten Euclidean dimensions;
this gives a basis of $2^{10}=1024$ matrices.
Of these $1024$ matrices, $496$ are antisymmetric (corresponding to $\so(32)$ realised by real roots $\alpha$ and the only relevant ones for spin $1/2$) and $528$ are symmetric. In order for $J=X\otimes \Gamma$ to be antisymmetric with respect to~\eqref{invariant}, the antisymmetric $\Gamma(\alpha)$ have to be paired with symmetric polarisation tensors $X(\alpha)$ and vice versa, see~\eqref{constsplit}. Symmetric matrices $\Gamma(\alpha)$ are obtained for 
$\alpha^2 \in 4 \mathbb{Z}$, so unlike spin-$\frac12$ they require null roots for their
realization. Altogether this leads to $51\,040= 496\times 55+ 528\times 45$ elements 
of $\so(288,32)$, where now also elements of the null root spaces are represented faithfully.

For the proof we have to verify that the basis dimensions are realised in the representation 
space, and this can be done by keeping a $\Gamma(\alpha)$ fixed and analysing all 
$X(\beta)$ such that $\Gamma(\beta) = \Gamma(\alpha)$ (there are infinitely many
such $\beta$'s for $\e$). Actually, it suffices to verify this property just for a smaller 
set of representative choices of $\Gamma(\alpha)$ as roots that are related by Weyl transformations do not have to be considered separately, as Weyl transformations 
rotate both the roots and the associated polarisation tensors. For real roots of $\e$ it is
in fact enough to do this for one simple root $\alpha_i$  because all (infinitely many)
real roots are Weyl conjugate.

This analysis can be done on a computer and we find that all the solutions 
from the constraint~\eqref{constraint} are realised in the representation space:
this means that the quotient algebra of the spin-$\frac{3}{2}$ representation is 
\begin{equation}
\q_{3/2}(K(\e))  \cong  \so(288,32) \, .
\end{equation} This result is an interesting outcome: the maximal 
\emph{compact} subalgebra $\ke$ gives rise to the \emph{non-compact} quotient 
$\so(288,32)$. This surprising feature is shared by all known spin $>\frac{1}{2}$ 
representations, and  has no analog in the case of finite-dimensional Lie algebras,
as will be further discussed below.

\subsubsection{Spin-$\frac{5}{2}$ and spin-$\frac{7}{2}$}

The very same procedure can be applied to the spin-$\frac52$ representation:
here, the invariant bilinear form is
\begin{equation}
(\phi,\psi) = \phi^\ind{ab} G_\ind{ac} G_\ind{bd} \psi^\ind{cd} ,
\end{equation}
which has signature $(1472,288)$.
Hence, the expected quotient algebra is $\mathfrak{so}(1472,288)$: again, 
a non-compact quotient. 
Following our general procedure we solve the constraint~\eqref{constraint} for $X$ 
(with the aid of the $\Gamma$'s basis), and the solution we find is a set of 1540 symmetric 
$X$ matrices associated to the antisymmetric $\Gamma$'s and 1485 antisymmetric
$X$'s associated to the symmetric $\Gamma$'s, which combine to a basis of $\so(1472,288)$.

Again, we now would have to check that the solutions 
from~\eqref{constraint} are all realised in the algebra. 
To be sure, we have so far not been able to generate a complete basis for the polarisation 
tensors in the algebra by iterating commutators, so we do not yet have a definite proof
whether all solutions of the constraint are actually realised in $K(E_{10})$ or not.
However, the number of independent elements in $\ke$ generated with our procedure
 is already so large that the resulting quotient algebra cannot be anything else but
 \begin{equation}
 \q_{5/2} (K(\e))  \cong  \so(1472,288)\,.
\end{equation}
where now also elements of root spaces associated to timelike imaginary roots
with $\alpha^2 = -2 $ are represented faithfully.
Finally, the conjectured quotient for the spin-$\frac72$ representation is
$\so(5568,1472)$ but to verify this claim would require even more extensive
checks than for spin-$\frac52$.

\subsection{General remarks on the quotients}

We found that the quotients for the spin-$\frac{3}{2}$ and spin-$\frac{5}{2}$ representations are non-compact, in marked contrast with the fact that they originate from a 
compact algebra, namely $\ke$. This is a special feature that can occur 
only for infinite-dimensional algebras.\footnote{We are indebted to R.~K\"ohl 
for discussions on this point.}

If $\g$ is a finite-dimensional simple and simply-laced algebra of rank at least 2, its maximal compact subalgebra $K(\g)$ is semi-simple, admitting no non-trivial solvable ideals.
Representations of $K(\g)$ are then completely reducible by the standard theory of semi-simple compact Lie algebras, and this applies in particular to the adjoint representation $K(\g)$ itself.
Therefore, given the existence of an ideal $\mathfrak{i}\subset K(\g)$, this is a representation of $K(\g)$ and, by complete reducibility, must split off as a direct summand of $K(\g)$, \emph{i.e.}~$K(\g) = \mathfrak{i} \oplus \mathfrak{q}$ as a sum of $K(\g)$ representations.
In this decomposition we have already indicated that $K(\g)/\mathfrak{i} \cong \mathfrak{q}$ is the remaining summand.
An explicit construction of $\mathfrak{q}$ as the orthogonal complement of $\mathfrak{i}$ in the (negative) definite invariant form on $K(\g)$ is mentioned at the end of section~\ref{sec:K}.
The bilinear form on $\mathfrak{q}$ as a subalgebra of $K(\g)$ is obtained by restriction and therefore preserves the definiteness property:
any subalgebra of $K(\g)$ must be compact.
An instance where this happens is the Cartan type $D_n$.
The split real form is $\g=\so(n,n)$ with $K(\g)=\so(n)\oplus \so(n)$, showing explicitly that $K(\g)$ is semi-simple, has ideals and that all subalgebras (= quotients) are compact.

For infinite-dimensional $K(\g)$, it is not known whether complete reducibility still holds and one can therefore not repeat the same argument as above.
Our investigations of the explicit unfaithful representations show that it can happen that an ideal $\mathfrak{i}$ does {\em not} split off as a direct summand and therefore the quotient Lie algebra $\mathfrak{q}=K(\g)/\mathfrak{i}$ need not be a subalgebra of $K(\g)$.
Whether or not $\mathfrak{q}$ admits an invariant bilinear form and whether this form is definite is a question that has then to be answered separately as this form 
does not follow from that on $K(\g)$ by restriction.
If $\mathfrak{i}$ should split off as a direct summand one could construct the quotient again as the orthogonal complement of $\mathfrak{i}$ as in the finite-dimensional case.
Below we shall use this construction to argue that for $K(E_9)$ the quotient cannot be a subalgebra (given by finite linear combinations of $K(\g)$ elements) but rather is distributional in nature. This shows that $\mathfrak{i}$ does not split off in that case.

This result also applies to $\ke$: for instance, the quotient $\so(288,32)$ is not contained in $\ke$ as a subalgebra.
Even though the \emph{image} of the $\ke$ generators in the \emph{unfaithful} representation behaves as $\so(288,32)$ matrices, the generators themselves in $\ke$ do not obey the $\so(288,32)$ algebra.
If one wanted to write combinations of $\ke$ that behave like $\so(288,32)$ in $\ke$, one would require non-convergent infinite sums of generators, whose
commutators could not be meaningfully evaluated. 

It is also worth noting that the $\ke$ ideals for increasing spin are not necessarily embedded into 
one another for increasing spin even though this might appear to be the case from the quotient algebras that we have determined above and the increasing faithfulness of the representation. 
Let us assume that $\mathfrak{i}_{3/2} \subset \mathfrak{i}_{1/2}$; then $\mathfrak{q}_{3/2}$ has to act on the spin-$\frac{1}{2}$ representation as well.\footnote{We thank G.~Bossard for pointing this out to us.}
But there is no way $\so(288,32)$ can act on the $32$ components of the 
Dirac representation spinor: this implies that the ideals are not contained in each other.
(This is in contrast with what happens for $K(E_9)$, where the ideals are in fact 
contained in each other, see~\cite{Kleinschmidt:2006dy} and below.)

A more manageable illustration of the fact that the quotient by the ideal 
$\mathfrak{i}$ is generally {\em not} a subalgebra of the infinite-dimensional
involutory algebra, unlike in the finite-dimensional case, is provided by the example of
the affine Kac--Moody algebra $E_9$ and its involutory subalgebra $K(E_9)$, and
the action of the latter on the fermions of maximal $D=2$ supergravity. Namely, as 
shown in~\cite{Nicolai:1991tt,Nicolai:2004nv,Kleinschmidt:2006dy}, 
to which we refer for further details, the involutory subalgebra $K(E_9)$ 
admits an explicit realization via the $E_9$ loop algebra; more precisely, it is
realised by elements
\begin{equation}\label{x(h)}
x(h) = \frac12 h^{IJ}(t) X^{IJ} + h^A(t) Y^A ,
\end{equation}
where $(X^{IJ}, Y^A)$ are the $120+128$ generating matrices of $E_8$ 
and $t\in\mathbb{C}$ is  the spectral parameter. The transformation (\ref{x(h)}) belongs to 
$K(E_9)$ Lie algebra if the (meromorphic) functions $\big(h^{IJ}(t),h^A(t)\big)$ 
obey the constraints
\begin{equation}
h^{IJ}\left(\frac1{t}\right) = h^{IJ}(t) , \quad
h^{A}\left(\frac1{t}\right) = - h^{A}(t) ,
\end{equation}
as well as the reality constraint exhibited in \cite{Nicolai:2004nv}.
For the Dirac (= spin-$\frac12$) representation the action of $x(h)$ on the
$(16 + 16)$ chiral  spinor components $\varepsilon^I_\pm$ is obtained by 
evaluation of the functions $\big(h^{IJ}(t),h^A(t)\big)$
at the distinguished points $t=\pm 1$ in the spectral parameter 
plane \cite{Nicolai:1991tt}\footnote{We note that each choice $t=\pm1$ yields an irreducible $16$-component chiral spinor of $K(E_9)$ but we treat the two choices together as they result from the 
decomposition of the 32 components of the $\ke$ Dirac spinor.} 
\begin{equation}
x(h)\cdot \varepsilon^I_+ = h^{IJ}(1) \,\varepsilon^J_+  , \quad
x(h)\cdot \varepsilon^I_- = h^{IJ}(-1) \,\varepsilon^J_- .
\end{equation}
Hence the quotient algebra is
\begin{equation}
{\mathfrak q}_{1/2} (K(E_9))  \,\cong\, \so(16)_+ \oplus \so(16)_- ,
\end{equation}
which is manifestly not contained in either $K(E_9)$ or $E_9$.
The spin-$\frac32$ representation consists of the 128 spinor components
$\chi^{\dot A}_{\pm}$ and the gravitino components $\psi^I_{2\pm}$ (we here adopt
the superconformal gauge, otherwise we would have an extra $(16+16)$ components 
$\psi^I_\pm$, see~\cite{Nicolai:2004nv,Kleinschmidt:2006dy}\footnote{With these extra components
  we would indeed recover the counting of the spin-$\frac32$ representation
  of $K(E_{10})$: $2\times (16 + 128 + 16) = 320$.}).
The action of $x(h)$ is now realised by  evaluation of $h$ and its first derivatives 
at $t = \pm 1$ \cite{Nicolai:2004nv}; more precisely,
\begin{align}
x(h)\cdot \psi^I_{2\pm} &= h^{IJ} (\pm 1) \, \psi^J_{2\pm} , \nonumber \\[2mm]
\label{ke9s32}
x(h)\cdot \chi^{\dot A}_\pm &= \frac14 h^{IJ}(\pm 1) \, \Gamma^{IJ}_{\dot A \dot B} 
\,\chi^{\dot B}_\pm 
\, \mp \, \frac12 (h')^A (\pm 1)\, \Gamma^I_{A\dot A} \psi^I_{2 \pm} .
\end{align}
The quotient algebra in the spin-$\frac32$ representation is thus
\begin{equation}
\label{ke9q32}
{\mathfrak q}_{3/2} (K(E_9))  \, \cong\,  \big[ \so(16)_+  \oplus \so(16)_-\big]
\,\oplus\,  \big[ \mathbb{R}^{128} _+ \oplus \mathbb{R}^{128}_- \big] ,
\end{equation}
which is a semi-direct sum and again not a subalgebra of $K(E_9)$. This is consistent with the
appearance of non-compact quotients in $\ke$ (which contains $K(E_9)$), as there is
no way to embed a translation group into a compact group (and indeed, the above 
transformations are contained in $\so(288,32)$). Furthermore, as
shown in~\cite{Kleinschmidt:2006dy}, there is an inclusion of ideals $\mathfrak{i}_{3/2}\subset \mathfrak{i}_{1/2}$ for $K(E_9)$ and this is consistent with the fact that~\eqref{ke9q32} can act on the $(16+16)$ spin-$\frac12$ components by having a trivial action of the translation part $\R^{128}_+\oplus \R^{128}_-$ on the spinor. We also note that the spin-$\frac32$ representation is indecomposable, in that the representation matrices of~\eqref{ke9s32} take on a triangular form
\begin{align}
x(h) \cdot \begin{pmatrix}\chi^{\dot{A}} \\[2mm]\psi_2^I \end{pmatrix}\sim \begin{pmatrix}
\so(16)_{\textrm{spin.}} & \R^{128}_{128\times 16} \\[2mm] 0 & \so(16)_{\textrm{fund.}} \end{pmatrix}\begin{pmatrix}\chi^{\dot{A}} \\[2mm]\psi_2^I \end{pmatrix}\,.
\end{align}
The indecomposability means that although $\psi^I_2$ is a subrepresentation of the 
spin-$\frac32$ representation it does not split off as a direct summand, showing 
again the failure of complete reducibility for $K(\g)$ in the Kac--Moody case. 

The difficulty of quotienting out  the ideal 
can be traced back to the distributional nature of the complement of the ideal in $K(E_9)$
for both examples. The ideals  $\mathfrak{i}_{1/2}$  and $\mathfrak{i}_{3/2}$,
respectively, are spanned by those elements $x(h)$ in (\ref{x(h)}) for which either
$h^{IJ}(\pm 1) = 0$, or $h^{IJ}(\pm 1) = 0$ {\em and} $(h')^A(\pm 1) = 0$. Consequently, the 
complement of either ideal would have to have $\delta$-function-like support
at $t=\pm 1$ in the spectral parameter plane, and thus cannot be represented in 
the form (\ref{x(h)}) with smooth functions $h$~\cite{Kleinschmidt:2006dy}. 
We anticipate that the difficulties for $K(E_{10})$ are of a similar, but more severe
nature, probably requiring a substantial generalisation of known concepts
of distribution theory, as it is not clear whether the formally divergent series 
can be handled by established tools of distribution theory, as was the case for $K(E_9)$.

\section{$\ke$ and Standard Model Fermions}

We now come to a tantalizing, but much more speculative feature of the finite-dimensional 
spin-$\frac32$ spinor representation of $\ke$, and a feature that may eventually allow to 
link up the abstract mathematical theory developed here to Standard Model physics. Namely,
for $\e$ it turns out that the spin-$\frac32$ representation, when viewed from the
point of view of four-dimensional physics, can be interpreted as the combination of
eight massive gravitinos and 48 spin-$\frac12$ matter fermions, a remarkable 
coincidence with the fact that the observed quarks and leptons in the Standard Model 
(including right-handed neutrinos) come in three families of 16 spin-$\frac12$ fermions:
$48 = 3 \times 16\,$!

This coincidence was already foreshadowed by the fact that maximal $N\!=\!8$ supergravity,
in four dimensions, after complete breaking of supersymmetry, is characterised by 
48 spin-$\frac{1}{2}$ fermions. As first stressed by Gell-Mann in~\cite{GellMann:1986ni}, 
this number matches that of the 48 fermions of the Standard Model if one
includes three right-handed neutrinos; moreover,  there 
is a way of putting the known quarks and leptons into representations of the 
residual symmetry group $\SU(3)\times\U(1)$  of $N\!=\!8$ supergravity remaining
after the gauge group $\SO(8)$ is broken, {\em provided} the supergravity SU(3) is 
identified with the diagonal subgroup of the color group SU(3)$_c$ and a putative 
family symmetry group SU(3)$_f$. However, there remained a mismatch 
in the  electric charge assignments by $\pm \frac16$ that was only recently fixed: namely,
as shown in~\cite{Meissner:2014joa}, the correct charges are obtained by deforming 
the subgroup $\U(1)$ in the way explained below that is not compatible with the 
supergravity theory itself. The required deformation is, however, 
contained in $\ke$~\cite{Kleinschmidt:2015sfa} acting on the spin-$\frac32$ representation
of $\ke$! The deformation of the $\U(1)$ subgroup of $\SU(3)\times\U(1)$ 
required to rectify the mismatch of electric charges is obtained by acting on the 
tri-spinors $\chi^{ijk}$ of $N=8$ supergravity which transform in the 
$\mathbf{56} \equiv \mathbf{8}\wedge\mathbf{8}\wedge\mathbf{8}$  of $\SU(8)$, 
with the 56-by-56 matrix  $\exp \left( \frac{1}6 \omega \mathcal{I}\right)$, where
\begin{equation}
\label{matI}
\mathcal{I} \coloneqq \frac{1}{2} \big(T \wedge \id \wedge \id + 
\id \wedge \, T \wedge \id + \id \wedge \id \wedge \,T + T \wedge T \wedge T \big) ;
\end{equation}
here the matrix $T$ is defined by
\begin{equation}\label{T}
T \coloneqq
\begin{pmatrix}
0 & 1 & 0 & 0 & 0 & 0 & 0 & 0 \\
-1 & 0 & 0 & 0 & 0 & 0 & 0 & 0 \\
0 & 0 & 0 & 1 & 0 & 0 & 0 & 0 \\
0 & 0 & -1 & 0 & 0 & 0 & 0 & 0 \\
0 & 0 & 0 & 0 & 0 & 1 & 0 & 0 \\
0 & 0 & 0 & 0 & -1 & 0 & 0 & 0 \\
0 & 0 & 0 & 0 & 0 & 0 & 0 & 1 \\
0 & 0 & 0 & 0 & 0 & 0 & -1 & 0
\end{pmatrix}
\; .
\end{equation}
This matrix belongs to $\so(8)$ and represents an imaginary unit, with $T^2=-\id$, which in 
turn implies $\mathcal{I}^2=-\id$. The presence of the triple product $T \wedge T \wedge T$ 
in $\mathcal{I}$ is the reason why $\mathcal{I}$ is \emph{not} an $\su(8)$ element, hence
does not belong to the R symmetry of $N\!=\!8$ supergravity (confirming the long known
fact that this theory by itself cannot account for the charge assignments of the Standard Model 
fermions). Enlarging the $\SU(8)$ symmetry to $\ke$, however,  
one can show~\cite{Kleinschmidt:2015sfa} that in fact $\mathcal{I}$ does
belong to the infinite-dimensional algebra $\ke$ or, to be completely precise,
to the associated quotient $\mathfrak{q}$. Importantly, it is not possible to construct 
the element~\eqref{matI} without the over-extended root of $\e$, the root system of 
any regular subalgebra such as $E_9$ being too restricted. This could make $\e$ 
a crucial ingredient in connecting supergravity to the Standard Model particles: 
no finite-dimensional or affine subalgebra can possibly achieve this.

Since $\mathcal{I}$ does not belong to the R symmetry SU(8), one can now ask which bigger
group is generated by commuting $\mathcal{I}$ with SU(8) in all possible ways in the 
${\bf 56}$ representation of SU(8). Indeed, by repeated commutation of $\mathcal{I}$ with 
elements of $\su(8)$, one ends up with the bigger algebra $\su(56)$, 
which should thus be viewed as a subalgebra of $\so(288,32)$. To show this
we take the standard $\su(8)$ basis in the fundamental representation, 
made of by 28 antisymmetric matrices (nothing but the $\so(8)$ basis) and 35 symmetric matrices multiplied by the imaginary unit. These matrices belong to the $\mathbf{8}$ representation; 
we pass to the $\mathbf{56}$ representation of $\su(8)$ by means of the antisymmetrised product used in~\eqref{matI}. Then we commute $\mathcal{I}$ with the 63 elements of the basis;
we collect the new independent elements from the commutators, and redo the procedure again until we stabilise the set of matrices.
Proceeding in this way, after four iterations we reach a stable set, \emph{i.e.}~no new elements are produced by the commutators.
This new set has dimension 3135 and corresponds to $\su(56)$, since it is made of by $56 \times 56$ anti-hermitian and traceless matrices.

Because $\mathcal{I}\in\ke$, the group $\SU(56)$ is therefore contained in the closure of SU(8)
and $\mathcal{I}$, and it is thus a subgroup of the vector spinor quotient $\Spin(288,32)/\Z_2$ (at the level of algebra, the quotient is $\so(288,32)$). Let us also mention that the very same calculation can be done with $\so(8)$: starting from the standard basis (28 antisymmetric matrices) in the $\mathbf{8}$ representation, we build the $\mathbf{56}$ representation and start to commute with $\mathcal{I}$.
Proceeding in this way, we reach 1540 independent antisymmetric $56 \times 56$ matrices
to obtain the desired $\so(56)$ basis.

Perhaps even more importantly, it has been shown in recent work~\cite{Meissner:2018gtx}
that actually the full (chiral) Standard Model group 
SU(3)$_c \times$ SU(2)$_w \times$ U(1)$_Y$ can be embedded into this
unfaithful representation of $\ke$, thus in particular undoing the descent to the 
diagonal subgroup SU(3) of the color and family groups, with a hypothetical
family symmetry SU(3)$_f$ that does not commute with the electroweak symmetries
(because the upper and lower components of the would-be electroweak doublets 
must be assigned to opposite representations of SU(3)$_f$ \cite{GellMann:1986ni}), 
That such an embedding is possible is not entirely unexpected in view of the fact that 
the above $\SU(56)$ acts {\em chirally} in four dimensions and 
via its subgroup SU(48) is large enough to contain 
the Standard Model  groups as subgroups (as well as a novel family symmetry). So the more
intricate open question and the true challenge is to understand whether and why this symmetry 
breaking might be preferred in $\ke$. We also reiterate that the eight massive gravitinos are
an integral part of the spin-$\frac32$ representation, and thus must likewise
be incorporated into the physics. Indeed, a  possible interpretation as (Planck mass)
Dark Matter candidates was suggested for them in~\cite{Meissner:2018cay},
together with possible experimental tests of this proposal. However,
the main challenge that remains is to see how $\ke$ can `unfold' 
in terms of bigger and increasingly less unfaithful $\ke$ representations to give rise
to actual space-time fermions, explaining how the above groups can be elevated
to {\em bona fide} gauge symmetries in space and time (we note that
explaining the emergence of bosonic space-time fields and symmetries 
from $E_{10}$ likewise remains an open problem). There remains a long way to go!

\vspace{1cm}\noindent
 {\bf Acknowledgments:}  H.~Nicolai would like to thank V.~Gritsenko and V.~Spiridonov
 for their hospitality in Dubna. His work has received funding from the European 
 Research Council (ERC) under the European Union's Horizon 2020 research and 
 innovation programme (grant agreement No 740209). We are grateful to G.~Bossard, 
 R.~K\"ohl and H.A.~Samtleben for comments and discussions.

\providecommand{\href}[2]{#2}\begingroup\raggedright\endgroup

\end{document}